\documentclass[journal]{IEEEtran}

\usepackage{amsmath,amssymb,amsfonts}
\usepackage{cite}
\usepackage{epsfig}
\usepackage{graphicx}
\usepackage{epstopdf}
\usepackage{algorithmic}
\usepackage{textcomp}
\usepackage{xcolor}
\usepackage{color,soul}
\usepackage[caption=false,font=footnotesize]{subfig}
\usepackage{xfrac}
\usepackage{balance}
\usepackage{url}
\usepackage{bbm}
\usepackage[pscoord]{eso-pic}

\def \R{\mathbb{R}}

\def \var{\operatorname{Var}}
\def \cov{\operatorname{Cov}}
\def \EER{\mathrm{EER}}
\def \corr{\operatorname{Corr}}
\def \E{\mathop{\mathbb{E}}}

\def \rhov{\boldsymbol{\mathrm{\rho}}}
\def \p{\boldsymbol{\mathrm{p}}}
\def \n{\boldsymbol{\mathrm{n}}}
\def \v{\boldsymbol{\mathrm{v}}}

\def \x{\boldsymbol{\mathrm{x}}}
\def \y{\boldsymbol{\mathrm{y}}}
\def \o{\boldsymbol{\mathrm{o}}}
\def \I{{\bf I}}
\def \H{{\bf H}}
\def \Hblur{{\bf H}_\textnormal{blur}}
\def \Hdeblur{{\bf H}_\textnormal{deblur}}
\def \Hdeblurhat{\hat{\H}_\textnormal{deblur}}
\def \S{{\bf S}}
\def \C{{\bf C}}

\def \G{{\bf G}}
\def \L{{\bf L}}
\def \e{{\bf e}}

\def\PatchDimInch{$\frac{2}{3}$-by-$\frac{2}{3}$ inch$^2$ }

\newcommand{\placetextbox}[3]{
	\setbox0=\hbox{#3}
	\AddToShipoutPictureFG*{
		\put(\LenToUnit{#1\paperwidth},\LenToUnit{#2\paperheight}){\vtop{{\null}\makebox[0pt][c]{#3}}}%
	}%
}%

\graphicspath{{./figs/}{./figs2/}{./}}

\begin{document}

\title{On Microstructure Estimation Using Flatbed Scanners for Paper Surface-Based Authentication}

\author{Runze Liu,~\IEEEmembership{Student Member,~IEEE and}
        Chau-Wai Wong,~\IEEEmembership{Member,~IEEE}%
\thanks{R. Liu and C.-W. Wong are with the Department of Electrical and Computer Engineering and the Forensic Sciences Cluster, NC State University, Raleigh, NC 27695, USA. (e-mail: rliu10@ncsu.edu; chauwai.wong@ncsu.edu.) \\
}
}

\maketitle

\begin{abstract}
Paper surfaces under the microscopic view are observed to be formed by intertwisted wood fibers. Such structures of paper surfaces are unique from one location to another and are almost impossible to duplicate. Previous work used microscopic surface normals to characterize such intrinsic structures as a ``fingerprint'' of paper for security and forensic applications.
In this work, we examine several key research questions of feature extraction in both scientific and engineering aspects to facilitate the deployment of paper surface-based authentication when flatbed scanners are used as the acquisition device.
We analytically show that, under the unique optical setup of flatbed scanners, the specular reflection does not play a role in norm map estimation.
We verify, using a larger dataset than prior work, that the scanner-acquired norm maps, although blurred, are consistent with those measured by confocal microscopes.
We confirm that, when choosing an authentication feature, high spatial-frequency subbands of the heightmap are more powerful than the norm map.
Finally, we show that it is possible to empirically calculate the physical dimensions of the paper patch needed to achieve a certain authentication performance in equal error rate (EER).
We analytically show that log(EER) is decreasing linearly in the edge length of a paper patch.
\end{abstract}


\vspace{-1mm}
\section{Introduction} \label{sec:intro}
\vspace{-1mm}
When viewed under a microscope, mundane-seeming paper surfaces come to life, and a maze of intertwisted wood fibers creates a complicated random jungle of structure\cite{paperproject,Authentication,buchanan2005forgery,clarkson-09,slava-12,slava-13,slava-14,wong2015icip,wifs15,wong2017,liu2018enhanced,kauba2016towards,schraml2017feasibility}. 
The unique microscopic structure of the paper surface is physically unclonable and may be considered as a ``fingerprint,'' which can be used for protecting valuable merchandise such as drugs and wines and important documents such as birth certificates and checks.
Two categories of methods have been used to capture such unique structures of paper surfaces for authentication, namely, the optical/visual feature approach and the physical feature approach. 

The optical/visual approach relies on the visual appearance of the paper surface or handcrafted features derived from the visual appearance for paper identification.
Buchanan et al.\cite{buchanan2005forgery} used a laser scanner to capture the reflected intensity due to a moving focused line that was shined on the paper surfaces, and used cross-correlation of digitized intensity fluctuations for identification. 
As a proof-of-concept effort for paper-based identification, lasers achieved good performance, however, they may be too expensive to be used in practical applications.
Beekhof et al.\cite{beekhof2008secure} used macrolens-aided mobile phones to capture images of the rough paper surfaces. 
Minimum reference distance decoding and reference list decoding were used for identification, with a huge reduction in complexity compared to classic minimum distance decoding while maintaining performance.
Sharma et al.\cite{sharma2011paperspeckle} used paper speckles, i.e., the dark and bright spots on paper when illuminated by light, as a fingerprint for the paper surface, where images of the paper surface were taken by a camera with the aid of a microscope with a built-in LED. 
The Gabor transform was applied to the captured image, and a binary image was obtained by using the complex phase of the Gabor transform and zero thresholding. 
The fractional hamming distance was used to compare different binary images.
Instead of analyzing the light reflected from the paper surface, Toreini et al.\cite{toreini2017texture} captured optical features of paper texture using the light transmitted through the paper, and had satisfying authentication performance. However, this method can only be applied to scenarios in which a sheet of paper is not glued to a surface and the paper is relatively transparent. For example, it is difficult to capture the transmissive light for a label stuck to a bottle or for stock paper packaging.
The aforementioned methods for identifying paper surfaces are based on the optical/visual features, while their underlying physical features, such as the orientation of a microscopic surface, have been shown to possess greater discriminative power\cite{clarkson-09, wong2017}.

The orientations of the microscopic surfaces of a paper patch may be quantified by the \textit{norm map}, a collection of uniformly spaced surface normals projected onto the $xy$-plane. 
Clarkson et al.\cite{clarkson-09} proposed a method for estimating a scaled version of the norm map of a paper patch by acquiring the paper in opposite orientations using a flatbed scanner, assuming light reflection is fully diffuse. 
Instead of using a bulky flatbed scanner, Wong et al.\cite{wong2017} used a mobile camera to take multiple photos from different perspectives of a paper patch, estimating the norm map with the diffuse reflection model \cite{book} and the camera's geometry \cite{hartley2003multiple}. The estimated norm map was also verified by ground truth, a norm map acquired by a confocal microscope.
Liu et al.\cite{liu2018enhanced} formulated two improved norm map estimators by taking into account the ambient light and the cameras' internal brightness and contrast adjustment processes. 
They also used estimated surface normals to reconstruct heightmaps (3D surfaces) of paper patches, and discovered that using the high spatial-frequency components of heightmaps as the authentication feature can achieve better performance than using the norm map.

\begin{figure}[!t]
\centering
  \vspace{-0mm}
  \hspace{-0mm}
  \includegraphics[width=\linewidth]{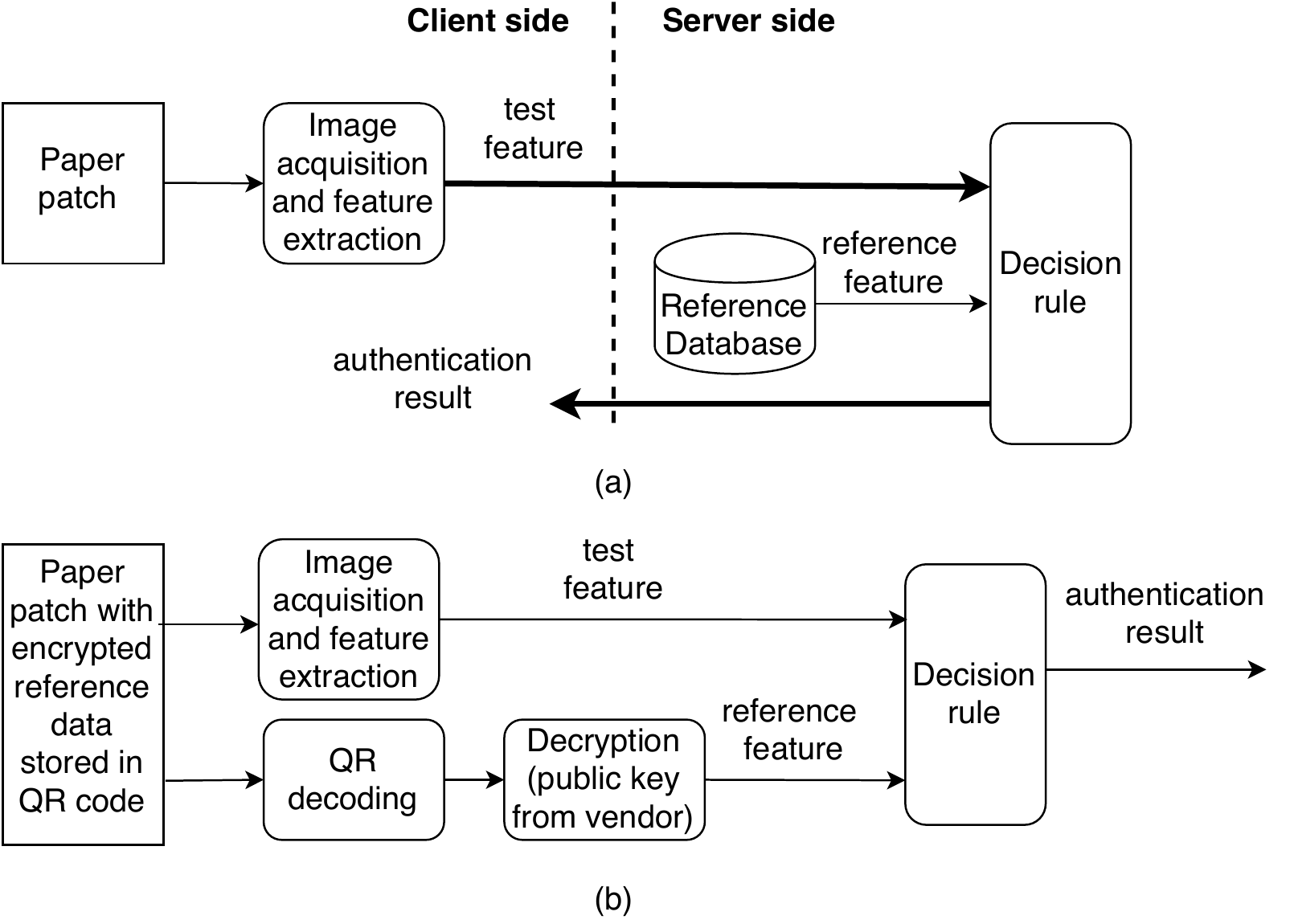}
  \caption{Examples of paper surface-based authentication systems: (a) a client--server model, and (b) a local model. The thick arrows are encrypted communication links and the normal arrows are local communication links. The diagrams focus on the verification stage. The reference data are stored in the reference database or the QR code at an earlier enrollment stage. 
  }
  \vspace{-3mm}
  \label{fig:high_level_diagram}
\end{figure}

Fig.~\ref{fig:high_level_diagram} demonstrates two potential designs of real-world paper surface-based authentication systems, namely, a client--server model and a local model.
The authentication systems, by designating a small paper-based surface area for the purpose of authentication, can be used for protecting merchandise and important documents.
For example, a customer can use a mobile camera to verify the authenticity of the packaging of drugs, and an institution can use a flatbed scanner to verify the authenticity of diplomas.
In the client--server model, a client with a mobile camera or flatbed scanner can acquire images of the paper patch, derive the test feature, and send the test feature to the server using a locally installed app.
The server will search in its database whether the test feature matches an existing reference feature upon receiving it from the client. If the reference feature ID is also provided together with the test feature, the server can directly access the reference feature and use it for comparison, which can save the feature retrieval time and increase authentication accuracy. The authentication result based on the matching outcome will be sent back to the client.  In the client--server model, the communication channel between the two parties is protected by cryptographic protocols, such as the transport layer security (TLS) to ensure trustworthiness. In the local model, the encrypted communication is not needed, but an additional QR code is used to store the reference feature protected by the public-key encryption. After decoding the QR code, the user will use the public key from the vendor to unlock the reference feature. The test feature will be compared with the reference feature to generate the authentication result. Although in this local model the reference feature may be exposed to an untrusted user that tries to tap into the memory to intercept the decrypted reference feature, the attacker still needs to forge a paper patch from which the intercepted feature can be derived, which is impossible because the microstructure is physically unclonable. 
Detailed use cases of the authentication systems and associated considerations are discussed in Section~\ref{sec:discussion}.

To facilitate the deployment of paper surface-based authentication, we examine four key research questions of feature extraction in both scientific and engineering aspects when flatbed scanners are used.
We chose to study scanners because they have a more controlled acquisition quality, which makes it easier for us to answer related research questions.
First, does ignoring the specular reflection have a destructive effect on authentication performance?
Prior approaches for estimating norm maps were based on the assumption that paper reflects light in a fully diffuse way\cite{clarkson-09,wong2015icip,wifs15,wong2017,liu2018enhanced}. 
In \cite{clarkson-09}, it was argued that the fully diffuse assumption largely holds, but without justification using experimental results or theoretical derivations. 
In \cite{wong2017}, the strengths of diffuse versus the specular components were estimated to be about six to one, but the specular was not compensated for in the norm map estimation.
Since the specular reflection could also be practically observed for paper surfaces even by the naked eye, it is interesting to investigate whether explicitly taking the specular reflection into the estimator design may improve accuracy.
Second, does the estimated normal vector resemble the real quantity with physical interpretations?
Prior work in \cite{wong2017} with a small dataset shows that norm maps acquired by scanners are consistent with those measured by confocal microscopes. In this work, we use a confocal dataset of one order of magnitude larger to obtain a more confident conclusion and extend the inquiry into the scanner's blurring effect.
Third, can feature engineering on the estimated normal vectors yield higher authentication performance?
The result in \cite{liu2018enhanced} demonstrated that the heightmap and its higher-frequency subbands as features outperform norm maps for mobile cameras. We investigate whether a similar conclusion can be drawn for flatbed scanners.
Fourth, we also study how the paper patch size affects authentication performance and investigate the justification for digitizing resolutions for paper patches.

We summarize the contributions of this paper compared to previous work \cite{ wong2017, liu2018enhanced} in both scientific and engineering aspects. The scientific contributions are as follows:
\begin{itemize}
    \item we prove mathematically that the effect of specular reflection can be ignored because of the unique imaging setup of flatbed scanners (but such a result is not true for the camera setup);
    \item we investigate quantitatively the performance drop due to the existence of the blurring effect in the scanner, and use a one-order-of-magnitude-larger dataset than that of \cite{wong2017} to confirm that scanners can capture meaningful physical quantities of paper surfaces.
\end{itemize}
The engineering contributions are as follows:
\begin{itemize}
    \item we justify and give a guide to the choices for paper patch size and resolution with mathematical and experimental results, and investigate quantitatively the performance drop due to spatial registration error;
    \item we confirm that using the heightmap as the feature proposed in \cite{liu2018enhanced} is also more discriminative than using the norm map for the case of the flatbed scanner.
\end{itemize}

The rest of the paper is organized as follows. 
In Section~\ref{sec:background}, we give some background reviews.
In Section~\ref{sec:deviations}, we analytically investigate the effect of specular reflection in the optical setup of flatbed scanners.
In Section~\ref{sec:consistency}, we investigate the consistency between estimated norm maps from scanners and the confocal microscope, with a focus on the blurring effect.
In Section~\ref{sec:3D}, we examine the performance of physical features, such as the heightmap and its subbands.
In Section~\ref{sec:discussion}, we investigate the digitizing resolution and the paper patch size needed for achieving a certain performance level.
Section~\ref{sec:discussion_application} discusses the potential applications.
Section~\ref{sec:conclusion} concludes the paper and discusses the future work.

\section{Background and Preliminaries} \label{sec:background}

Symbol conventions are as follows. Nonitalic bold lower cases of letters denote column vectors. 
For example, $\n=(n_x, n_y, n_z)^T$ is a column vector.
Nonitalic bold upper cases of letters denote matrices.

\subsection{Difference-of-Gaussians (DoG) Representation} \label{subsec:dog}
In DoG representation\cite{lowe2004distinctive,lindeberg1994scale}, the $n$th level subband is obtained by taking differences of the Gaussian-blurred matrix of numbers as follows:
 \begin{equation}
\L_n = \G_n - \G_{n+1}, \ n = 1,...,N
\end{equation}
where $\G_1$ is defined to be the original matrix, $\G_{N+1} = 0$, and $\G_n$, $n = 2,...,N$, 
is the result of blurring the original matrix by a Gaussian filter with standard deviation $\sigma^{n-1}$, where $\sigma > 1$. 
The DoG representation of a matrix allows us to investigate the different spatial-frequency subbands of the matrix, as shown in Section~\ref{subsec:heightmap_subbands} and the supplementary document.

\subsection{Generalized Light Reflection Model}
Fig.~\ref{fig:n_v} illustrates a microscopic portion of a paper surface containing small surfaces that usually orient differently than the macroscopic paper surface. 
Picking an arbitrary location $\p \in \R^2$ on the surface and assuming both diffuse and specular reflection types,
the perceived intensity $l_r$ for a sensor or an eye at a fixed distance away from $\p$ may be written as the following generalized light reflection model, i.e., the Phong shading model without the ambient light \cite{book}:
\begin{figure}[!t]
\vspace{-2mm}
\begin{minipage}[b]{0.98\linewidth}
  \centering
  \centerline{\includegraphics[width=7.0cm]{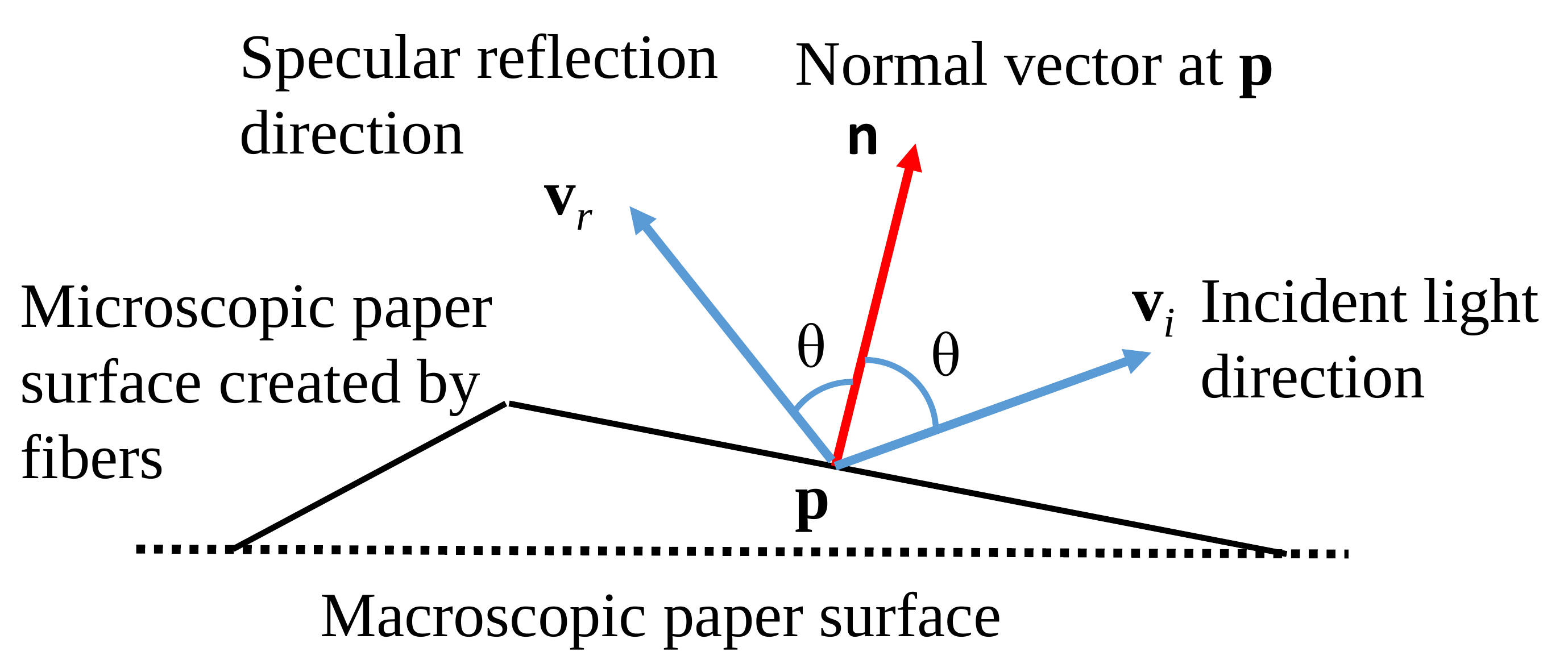}}
\end{minipage}
\vspace{-0.2cm}
\caption{A microscopic view of a paper surface with annotated quantities related to light reflection at location $\p$. The vectors are all unit vectors.}
\label{fig:n_v}
\end{figure}
\begin{equation}
l_r = \frac{l}{||\o-\p||^2} \Big\{ w_d  \cdot (\n^T\v_i)^+ + w_s  \cdot (\v_c^T \v_r)^{k_e} \Big\},
\label{sp}
\end{equation}
\noindent{}where $\n=(n_x, n_y, n_z)^T$ is the microscopic normal direction of the paper surface at location $\p$, $\o=(o_x, o_y, o_z)^T$ is the position of the light source,  $\v_i = (\o-\p) / || \o - \p ||$ is the incident light direction, $l$ is the strength of the light,
$1/||\o-\p||^2$ is a light-strength discounting factor as the received energy per unit area from a point light source, which is inversely proportional to the squared distance.
$x^+ = \max(0,x)$, and
$k_e>0$ controls the gloss level of the surface.
$w_d$ and $w_s$ are the weights for diffuse and specular components, taking into account the effect of a constant surface albedo and other scaling factors.
$\v_c$ is the camera's/sensor's direction, and $\v_r$ is the specular reflection direction which can be written in terms of the incident light direction $\v_i$ and the normal vector $\n$, i.e., $\v_r = (2\n\n^T-\I)\v_i$, where $\I$ is the identity matrix.
All $\n$, $\v_i$, $\v_c$, and $\v_r$ are unit-length column vectors.

\subsection{Norm Map Estimation Using Photometric Stereo} \label{subsec:clarkson}
A surface normal is a vector perpendicular to the tangent plane at a location of the surface, and a normal vector field is a collection of 3D surface normals over a 2D grid. 
A norm map is the normal vector field projected onto the $xy$-plane, which is a 2D vector field.
The norm map has been shown to be a powerful discriminative feature for paper surfaces\cite{clarkson-09,wong2017,liu2018enhanced}.

The state-of-the-art method for estimating norm maps of paper surfaces using commodity flatbed scanners\cite{clarkson-09,wong2015icip,wong2017,liu2018enhanced,wifs15} is described as follows. 
\begin{figure}[!t]
  \centering
  \vspace{-2mm}
  \includegraphics[width=0.8\linewidth]{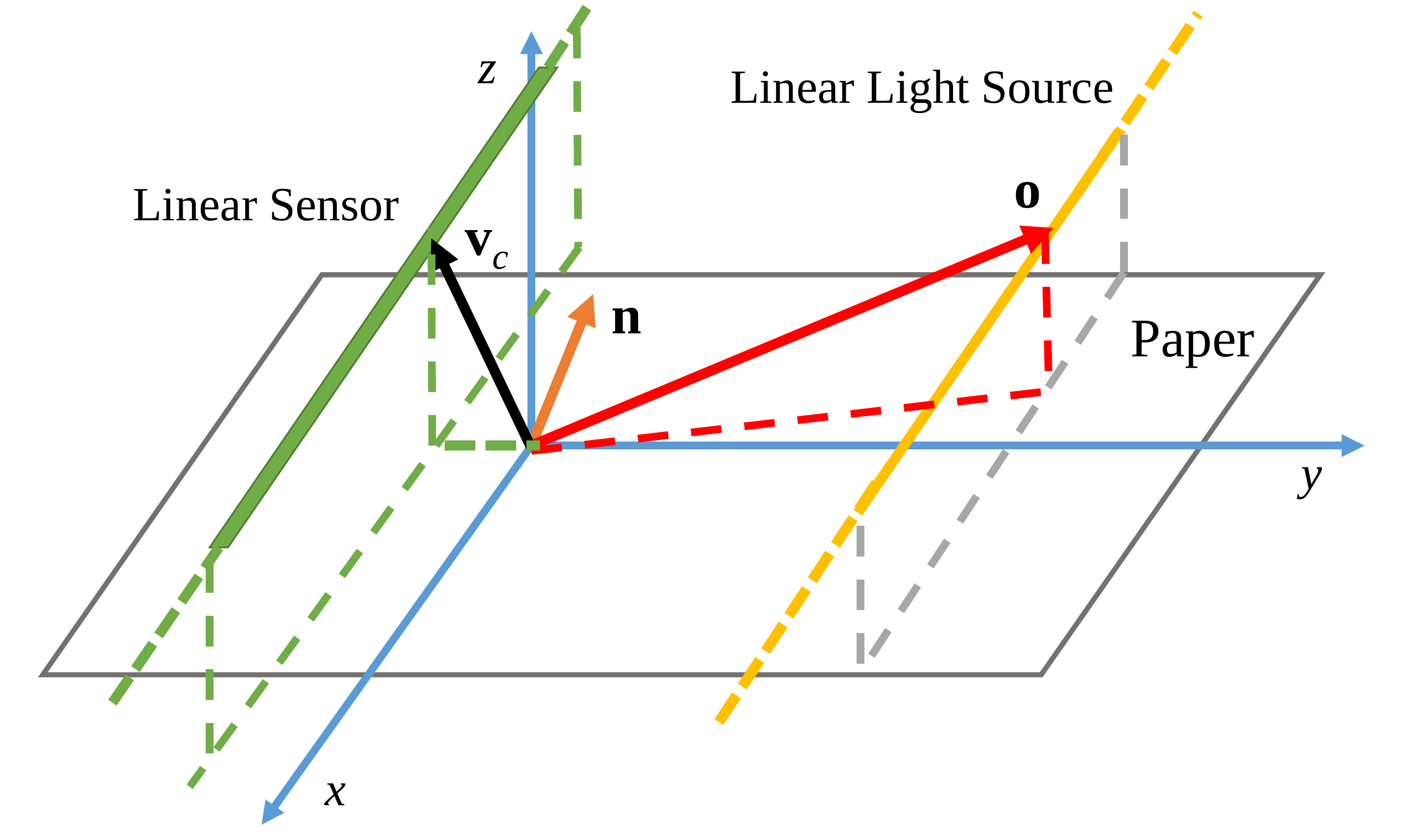}
  \vspace{-2mm}
  \caption{Configuration of the optical system of a flatbed scanner for scanning a paper sheet. The point of interest is located at the origin. The microscopic surface normal, $\n$, the camera/sensor direction, $\v_c$, and the location of one point on the linear light, $\o$, are shown.}
  \label{fig:optics}
\end{figure}
We assume the paper to be scanned is placed on the $xy$-plane passing through the origin as shown in Fig.~\ref{fig:optics}. 
Without loss of generality, we assume that the point of interest is located  at the origin. A linear light source is positioned in parallel with the  $x$-axis and moves along the $y$-axis. 
We denote a specific location on the linear light source as $\o$ and the incident light direction is therefore $\v_i = (o_x,o_y,o_z)^T/||(o_x,o_y,o_z)||$. 
Since the light source is very close to the paper surface, the linear light source appears to a point on the paper infinitely long in the $x$-direction.

Under the fully diffuse model, the intensity $I$ of the reflected light of the point placed at the origin under the linear light of a flatbed scanner is a superposition of all rays diffusely reflected and originating from the light source located at $\o = (o_x,o_y,o_z)^T$  for $o_x \in [-a,b]$:
\begin{equation}
\begin{split}
I & =  \int_{-a}^{b} l_r \ do_x \approx l\cdot w_d \int_{-a}^{a}  \n^T\frac{(o_x,o_y,o_z)^T}{||(o_x,o_y,o_z)||^3} \  do_x ,
\end{split}
\label{eq1}
\end{equation}
\noindent{}where $-a$ and $b$ are the $x$-coordinates of the two ends of the linear light source and we assume $0<a<b$ without loss of generality. The approximation in \eqref{eq1} makes use of the fact that the intensity of the point of interest contributed by the far portion $o_x \in (a,b]$ of the linear light source is very small, 
namely, $\int_a^b \n^T \o/||\o||^3 do_x \approx 0$.

In \cite{clarkson-09, wong2015icip,wong2017,liu2018enhanced,wifs15}, images acquired using a scanner from two opposite directions are used to estimate the $x$- or $y$-components of a norm map.
Two images, $I_{0^\circ}$ and $I_{{180}^\circ}$, are obtained when the paper is orientated at $0^\circ$ and $180^\circ$ on the $xy$-plane when being scanned. 
For a pixel of interest on the paper surface, the normal vector is $\n$, and a specific location on the light source is $\o = (o_x,o_y,o_z)^T$. 
When scanning the paper at $180^\circ$, it is equivalent that for the pixel of interest, the normal vector remains the same, while flipping the light’s $y$-coordinate, 
namely, changing the specific location on the light source into $\o' = (o_x,-o_y,o_z)^T$.\footnote{Note that this equivalence by flipping the $y$-coordinate of the light is only valid for the fully diffuse model. In Section III, which incorporates the specular component, we do not use this equivalence.}
Their difference, $I_{0^\circ} - I_{{180}^\circ}$, can be shown to be in proportion to the $y$-component of the norm map, $n_y$, and therefore can be used as an estimator for $n_y$ \cite{clarkson-09}:
\begin{equation}
\begin{split}
 I_{0^\circ} - I_{{180}^\circ}
 = l \cdot w_d \int_{-a}^{a} \n^T \frac{\o-\o'}{||(o_x,o_y,o_z)||^3} \ do_x  
     = s \, n_y ,
\end{split}
\label{eq2}
\end{equation}
\normalsize
\noindent{}where $s = 2l \cdot w_d o_y \int_{-a}^{a} ||\o||^{-3} do_x$ is a constant.
The $x$-component of the normal vector, $n_x$, can be estimated similarly using $I_{90^\circ} - I_{{270}^\circ}$. 

\section{Cancellation of Specular Components Under Flatbed Scanner Geometry } \label{sec:deviations}
\begin{figure}[!t]
\centering
  \vspace{-0mm}
  \hspace{-0mm}
  \includegraphics[width=\linewidth]{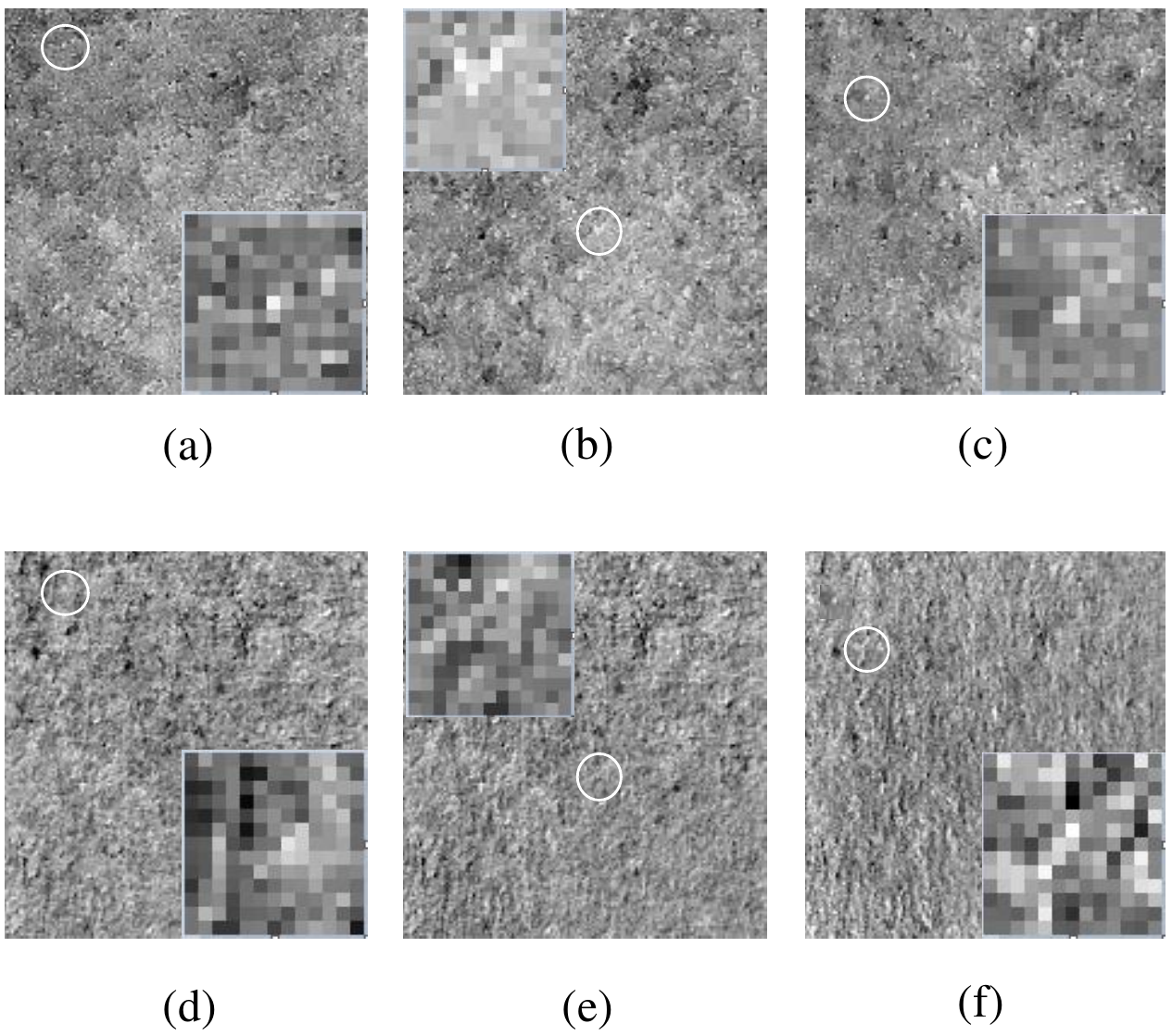}
  \caption{(a)--(c) Photos of a paper patch captured by a mobile camera from different angles with flashlight. (d)--(f) Synthetic images that consider only the diffuse reflection. 
The real photos have high-contrast spots that may be caused by specular reflection, whereas their contrast in respective synthetic images is much lower. 
Vertical paired images are to be compared, with circles highlighting collocated spots for visual comparison. The zoomed-in versions in the circled areas are put in the corners of the images.
(All pictures have undergone perspective transform, detrending, and contrast enhancement to better illustrate the idea.)
  }
  \vspace{-3mm}
  \label{fig:contrast_imgs}
\end{figure}

The state-of-the-art norm map estimation method \cite{clarkson-09,wong2017} reviewed in Section \ref{subsec:clarkson} assumes that paper surfaces reflect light in a fully diffuse way. 
However, if one observes carefully a paper patch at a close distance under a strong light while constantly changing the observation angle, he/she may observe some discrete spots with significant intensity fluctuation. 
These discrete spots are not fully diffuse, since perceived intensity due to diffuse reflected light should not depend on the location of the eye/sensor.
For a spot dominated by the specular reflection, the perceived intensity could be much stronger or weaker than its neighboring spots dominated by the diffuse reflection. This is because the intensity given by the specular reflection has a different cause that depends on the angle between the directions of the eye and the reflected light, namely, $\arccos(\v_c^T \v_r)$. 
For these spots with a specular reflection component, the estimation of the normal vector may be very different from the true value if the specular component is neglected.
To demonstrate this phenomenon, we contrast in Fig.~\ref{fig:contrast_imgs} real photos captured by a mobile camera and their corresponding synthesized versions, by only considering the diffuse component.
The photos were captured in different camera orientations with different incident light directions.
The synthesized versions were generated by first estimating the normal vector field, assuming the fully diffuse Model 2 proposed in \cite{liu2018enhanced}, and then rendering diffuse reflection images. 
It is revealed in Fig.~\ref{fig:contrast_imgs} that the real photos in the first row have more highlights than the synthesized images in the second row, which could be due to the specular reflection.
We circled some locations of high contrast in real photos that are surrounded by dark pixels. 
The corresponding locations in synthesized images do not have such high contrast.

We have demonstrated that, in general geometric setups for capturing paper surfaces, such as using cameras, there will be high-contrast spots in the captured images due to the specular reflection component.
Blindly ignoring specular reflections in modeling and estimation may lead to imprecise norm map estimates. 
Next, we show analytically that, for the flatbed scanner geometry, the image subtraction approach remains a precise estimator even if specular reflection is taken into consideration.
Using the generalized light reflection model \eqref{sp} that contains the specular reflection term, the reflected intensity under a scanner's linear light becomes:
\begin{equation}
\begin{split}
I & = \int_{-a}^{a} l_r do_x  = l \int_{-a}^{a} \big( w_d\n^T\v_i + w_s\v_c^T\v_r \big)\frac{1}{||\o||^2} \ do_x \\
 & = l \int_{-a}^{a} \Big( w_d\n^T + w_s\v_c^T(2\n\n^T-\I) \Big)\v_i\frac{1}{||\o||^2} \ do_x.
\end{split}
\end{equation}
\noindent{}Note that we set $(\n^T \v_i)^+ = \n^T \v_i$ when invoking \eqref{sp} since the angle between $\n$ and $\v_i$ are rarely greater than $90^\circ$. 
We set $k_e = 1$ to capture the dominating linear relationship while ignoring the higher-order terms for analytic tractability.

When scanning the paper in two opposite directions, a more natural and direct modeling approach is not to flip the light's $y$-coordinate as proposed in \cite{clarkson-09} and reviewed in Section \ref{subsec:clarkson} of this paper; 
instead, following the illustration of Fig.~\ref{fig:optics}, we should capture the $180^\circ$ rotation operation in the $xy$-plane resulting in $\n' = (-n_x,-n_y,n_z)$ while leaving the incident light direction $\v_i$ and the camera direction $\v_c$ unchanged. 
Following the traditional procedure of subtracting one scanned image from another, we obtain: 
\begin{equation}
\begin{split}
&I_{0^\circ} - I_{{180}^\circ} = sn_y  \\
&+ 2l \int_{-a}^{a} \bigg( w_s\v_c^T(\n\n^T-\n'\n'^T)\v_i \bigg)\frac{1}{||\o||^2} \ do_x.
\end{split}
\label{eq4}
\end{equation}

\begin{figure}[!t]
\centering
  \vspace{-0mm}

  \hspace{-4mm}
  {\includegraphics[height=3cm]{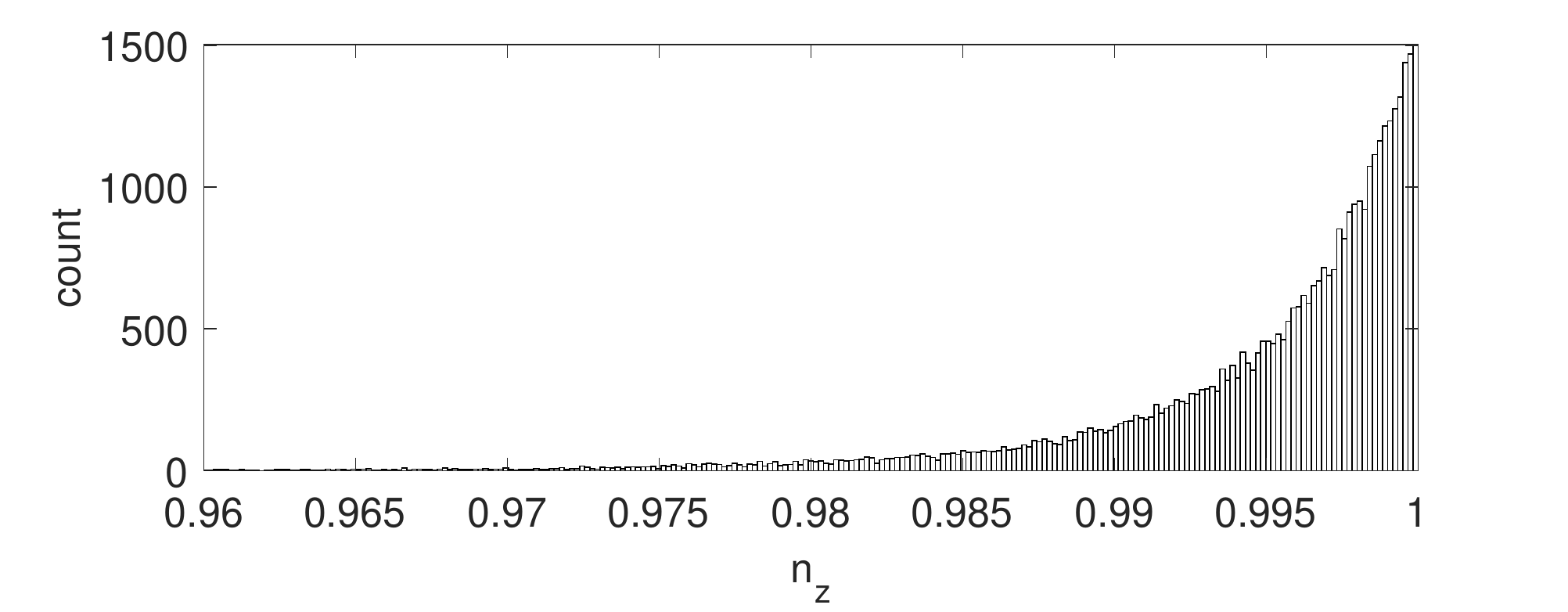}}
  \hspace{3mm}
  \caption{A histogram for the $z$-component of the normal vector field of a $2/3$-by-$2/3$ inch$^2$ paper patch from confocal laser scanning microscope Keyence VKx1100 digitized at a spatial resolution of $5.38 \ \mu $m.
	}
  \vspace{-3mm}
  \label{fig:nz_dist}
\end{figure}

\noindent{}The $x$-component of camera direction $v_{cx} = 0$ since the camera/sensor in the scanner catches the light that is parallel to the $yz$-plane, and $n_z\approx1$ since normal vectors are close to pointing straight up, as is revealed by Fig.~\ref{fig:nz_dist}---a histogram for $n_z$ obtained from measurements using a confocal microscope.
Substituting $\v_c = (v_{cx},v_{cy},v_{cz})^T$, $\v_i = \o/||\o||$ and $\n\n^T-\n'\n'^T = 
\scriptsize
\begin{bmatrix}
0 & 0 & 2n_xn_z \\
0 & 0 & 2n_yn_z \\
2n_xn_z & 2n_yn_z & 0
\end{bmatrix}
$ into \eqref{eq4},
we obtain:

\begin{subequations}
\begin{align}
\begin{split}
&I_{0^\circ} - I_{{180}^\circ} = sn_y + 4l \int_{-a}^{a} w_s(v_{cz}n_xn_z,v_{cz}n_yn_z,\\
& v_{cx}n_xn_z+v_{cy}n_yn_z) 
(o_x, o_y, o_z)^T ||\o||^{-3} \ do_x \label{eq3a} \\
\end{split}
\\
& = sn_y + 2s' n_z \Big\{n_y\Big[ v_{cz} + v_{cy} o_z / o_y \Big] + n_x v_{cx} o_z / o_y \Big\} \label{eq3b} \\
& \approx \left[ s+  2(v_{cz} +v_{cy} o_z / o_y) s' \right]n_y
\label{eq3}
\end{align}

\end{subequations}
\noindent{}where $s' = 2l \cdot w_s o_y\int_{-a}^{a}{||\o||^{-3}}do_x$. 
We followed the procedure outlined in \cite{wong2017} to generate normal vectors from the heightmap acquired by a confocal microscope. 
Note that $o_z$ and $o_y$ are device-specific constants since the distance from the light source to the point being captured in the $xz$-plane is fixed by the design of the scanner geometry.
The final result in \eqref{eq3} reveals that even though the specular reflection is taken into account, the traditional estimator is still linear in $n_y$ due to the unique imaging setup by flatbed scanners. 
This would not be possible if $v_{cx}$ were not zero, since both $n_x$ and $o_z$ are usually nonzero. 

Note that the result that the specular component does not play a role is largely due to the approximately symmetric integration bound from $-a$ to $a$, as demonstrated in \eqref{eq1}, which in turn is guaranteed by the fact that the linear light is very close to the paper to be scanned in the $z$-direction.
The result we obtained in this section does not apply to more general geometric setups, such as using mobile cameras discussed in other literature \cite{slava-12,slava-13,slava-14,wifs15,wong2017,liu2018enhanced}.
This result also justifies the use of a flatbed scanner to obtain norm maps for surfaces other than paper that contain stronger specular components.

\section{Scanner and Confocal Consistency} \label{sec:consistency} 

A preliminary study was reported in Section VII.C of \cite{wong2017} examining whether the norm map estimated from scanner-acquired images is consistent with the ground truth, i.e., with the norm map measured by the confocal microscope.
The overall correlation between the scanner estimates and the reference was $0.28$ (we reproduced this number in Table \ref{tab:corr_summary} for easy reference and comparison), indicating that the estimation, even though not very precise,  was indeed related to the ground truth. 
However, in \cite{wong2017}, only one physical paper patch was investigated. 
In the present paper, we extend the inquiry of \cite{wong2017} by using a confocal-collected dataset of one order of magnitude larger, and we investigate the blurring issue, aiming to confirm with higher confidence the hypothesis that the scanner-estimated norm maps are meaningful physical quantities, thereby yielding a better understanding of the characteristics of the scanner-estimated norm maps.

\begin{table}[!t]
  \caption{Comparison of Performance of Various Features When Test Data From Scanner Correctly Match with Reference Data From Confocal Microscope\vspace{-2mm}}
  \label{tab:corr_summary}
  \centering
  \scalebox{1.0}{
  \begin{tabular}{lc}
\hline \hline
\textbf{Feature} & \textbf{Correlation} \\ \hline
\emph{Norm Map Based:}\\ \hline
Raw (dataset of \cite{wong2017}) & $0.28$ \\
Raw (new dataset) & $0.357$ ($x$), $0.301$ ($y$)\\ 
Deblurred (new dataset) & $0.442$ ($x$), $0.396$ ($y$) \\ 

\hline
\emph{Heightmap Based:} \\
\hline
Reconstructed heightmap & $0.358$ \\
Detrended reconstructed heightmap & $0.499$ \\
Third-highest spatial-frequency subband & $0.714$ \\
\hline
\hline
\end{tabular}
}
\vspace{-1mm}
\end{table}

\subsection{Dataset Collection}
In this paper, we created a new dataset of paper surfaces that was made publicly available on the authors' websites.
We collected data for $9$ different paper patches of size $\frac{2}{3}$-by-$\frac{2}{3}$ inch$^2$ using flatbed scanners and a confocal microscope.
The patches are from the same sheet of ordinary office printing paper. 
This is a more difficult case than the case in which paper patches are obtained from different sheets of printing papers, because the paper patches from the same sheet exhibit less variations due to the same manufacturing condition, time, and raw materials used.
The papers with printing are not considered since we aim to derive the intrinsic physical features caused by the intertwisted wood fibers on the paper surface.
Four out of nine paper patches were stuck to a microscope glass slide to create a rigid and consistently flat surface. A card stock was put between the paper and the glass slide to block any light from the backside of the paper. The other five paper patches were not stuck to anything.
These two different setups mimic the conditions of patches in real-world scenarios.

Data related to the flatbed scanner include scanner-acquired images. 
For image acquisition, we used a Canon CanoScan LiDE 110 flatbed scanner to acquire each patch from four orientations, i.e., $0^\circ$, $90^\circ$, $180^\circ$ and $270^\circ$, and repeated the processes three times for each physical patch in order to obtain three norm maps. 
The norm maps were estimated by taking the difference of images scanned in opposite directions, a method that is based on the fully diffuse model, since we have analytically proven in Section \ref{sec:deviations} that the specular component can be neglected in the optical setup of a scanner.
Then we repeated the image acquisition process by using two other consumer-grade flatbed scanners that are the most popular on Amazon.com as of the summer of 2019: CanoScan Lide 300, and Epson Perfection V39. 
Using the three scanners, we obtained a total of nine norm maps for each paper patch.
We resized the acquired patch images to $200$-by-$200$ pixels.
Data related to the confocal microscope include heightmaps of paper surfaces and norm maps derived from heightmaps that are accurate enough to be considered as ground truth.
We used a Keyence VKx1100 confocal microscope with a $404$~nm violet laser source to obtain heightmaps of paper patches. 
We followed the procedure in \cite{wong2017} to derive a $200$-by-$200$ norm map from the heightmap for each paper patch: 
we estimated the normal vector for a pixel of interest by fitting a plane to the corresponding height values located in the $z$-direction.
The resolution in the $z$-direction of the heightmap used in this data acquisition was $0.1$~nm, which is much higher than $6$~$\mu$m used in \cite{wong2017} and can therefore provide more accurate aggregated results for confocal-generated norm maps. 
Due to the optical principles of confocal microscopy, the confocal norm map is accurate enough to be considered as the ground truth.

\begin{figure}[!tb]
	\centering
	\vspace{-3mm}
	\hspace{-1mm}
	\subfloat[]{\includegraphics[width=0.48\linewidth]{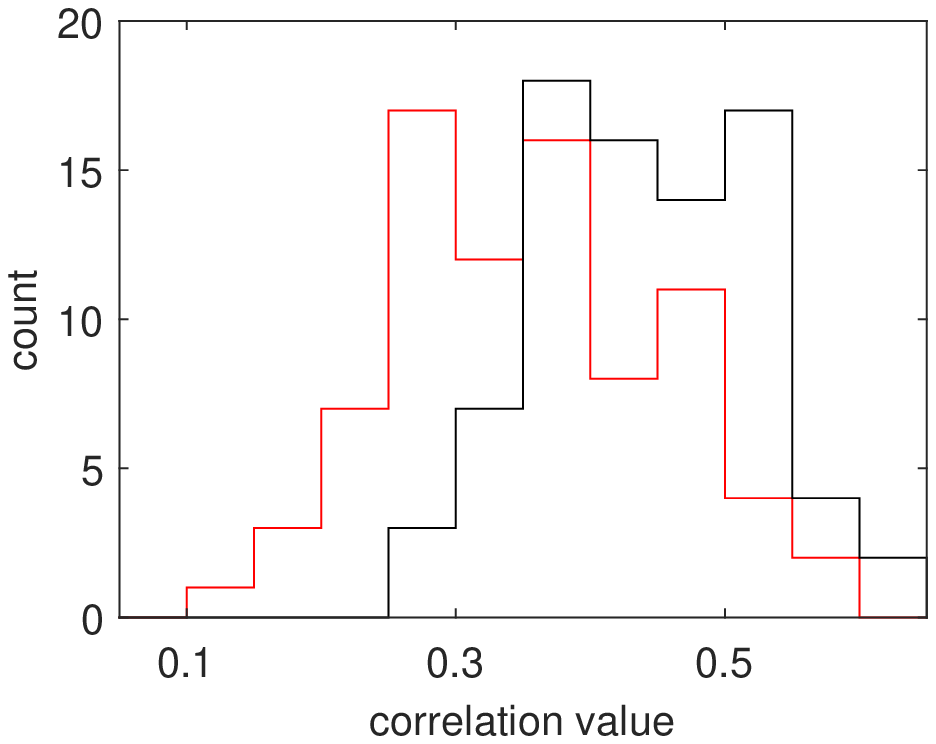}}
	\hspace{2mm}
	\subfloat[]{\includegraphics[width=0.48\linewidth]{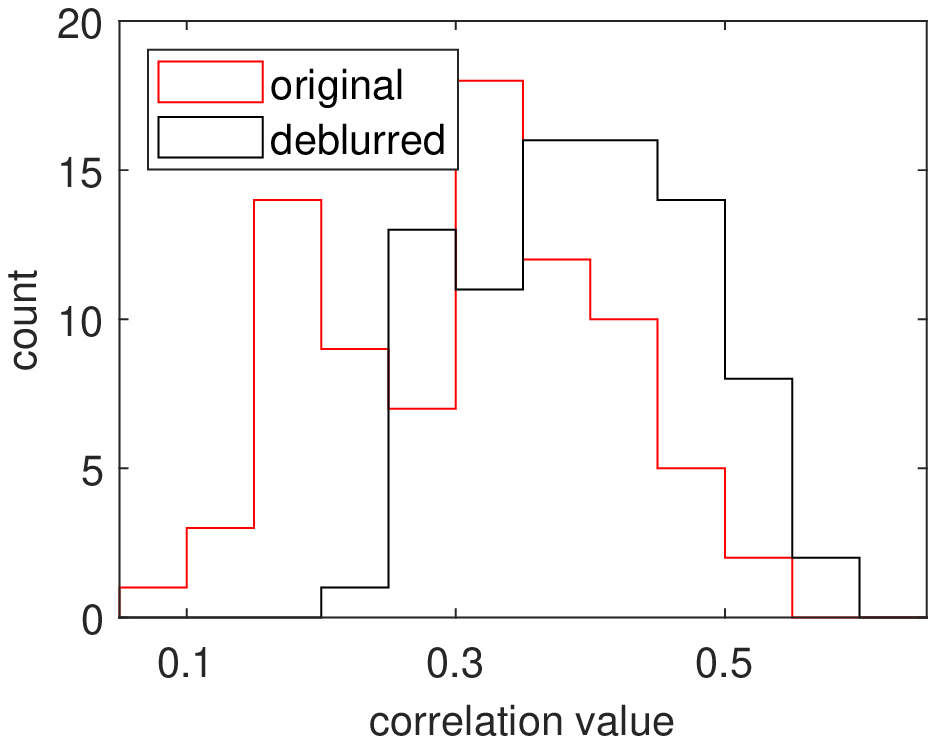}}
	
	\caption{Histograms of correlation values between the (a) $x$- or (b) $y$-component of norm maps estimated from scanner and confocal measurements. The averaged correlation increased from $0.357$ to $0.442$ for the $x$-component and from $0.301$ to $0.396$ for the $y$-component after deblurring.}
	
	\label{fig:nm_corr_dist}
\end{figure}

\subsection{Initial Consistency Verification}
We evaluated the consistency of the scanner-estimated norm maps to the confocal measurements on the newly collected dataset following the same procedure as in \cite{wong2017}.
For each paper patch, we calculated the correlation between the $x$-/$y$-component of the nine norm maps obtained from the scanners, and the ground-truth norm map from the confocal microscope. 
Histograms of the correlation values are shown using the ``original'' legend of Fig.~\ref{fig:nm_corr_dist}. 
The averaged correlation is $0.357$ for the $x$-component of the norm map and $0.301$ for the $y$-component, as summarized in Table~\ref{tab:corr_summary}, 
with the sample standard deviation $0.10$ and $0.11$, respectively. 
The averaged correlation values are close to the result, $0.28$, reported in \cite{wong2017}.
Our experimental results, from using nine different paper patches, confirm with higher confidence that the scanner-estimated norm maps are meaningful physical quantities.

\subsection{Consistency Verification by Compensating Blurring}
Although the previous subsection confirms that scanner norm maps are meaningful estimates of physical quantities, the correlation of slightly greater than $0.3$ implies that there are still non-negligible factors contributing to the inconsistency.
One such factor may be spatial blurring.
In this subsection, we investigate the blurring effect due to the imaging pipeline of flatbed scanners on the accuracy of the estimated norm maps.
Images captured by flatbed scanners may be blurred due to being out of focus, or to sensor/light/scanning platform motion.
Norm maps derived from blurred scanned images will therefore be a blurred version of the ground-truth norm maps.
Below, we examine whether deblurring is possible with the help of confocal norm maps and we investigate the characteristics of blurring filters.

\subsubsection{Deblurred Norm Map}
We explore using confocal norm maps to assist the deblurring process and we evaluate the quality of deblurred norm maps.
We denote the norm map from the confocal measurement as $\C$, and the norm map estimated by subtracting the two images scanned in opposite directions as $\S$.
We model the relation between the ground truth $\C$ and the scanner norm map considered to be blurred using the linear model:
\begin{equation}
  \C = \Hdeblur * \S + \e,
\label{eq:deblur_model}
\end{equation}
where $\Hdeblur$ is a linear spatial invariant (LSI) deblurring filter, $\e$ is an error term, and $*$ is the 2D convolution operator. 
We create separate models for $x$- and $y$-components of a norm map and for each paper patch.
Regarding the size of the deblurring filter, we empirically set the dimensions such that the pixels with significant contributions to the convolutional result will be retained.
Specifically, we use an oversized filter, i.e, $25$-by-$25$, to preliminarily estimate filter coefficients when the filter dimensions are not significantly constrained.
Since it is a deblurring filter, the coefficient of the pixel in the center must dominate in magnitude when compared to other pixels.
We observe that most coefficients with magnitude greater than $10\%$ of that of the centering pixel are located in the centering $7$-by-$7$ area.
Hence, we will use $7$-by-$7$ as the size by which to formally estimate the deblurring filters as follows.

To avoid model overfitting, we estimate the deblurring filter $\Hdeblur$ using cross-validation with the cost function in the ridge regression form:  
\begin{equation}
\min_{\Hdeblur} ||\C - \Hdeblur * \S||_F^2 + \lambda ||\Hdeblur||_F^2 ,
\end{equation} 
where $||\cdot||_F$ is the Frobenius norm and $\lambda$ is a regularization parameter controlling model complexity.
With a norm map of size $200$-by-$200$, and the filter size of $7$-by-$7$, there are $34,596$ data points to solve for $\Hdeblur$.
We first use 10-fold cross-validation to find the regularization parameter that minimizes the cross-validation error.
We then apply one standard error rule to choose an updated regularization parameter that corresponds to the most parsimonious model and use the coefficients at this time as the final estimate for the deblurring filter, $\Hdeblurhat$.

We use the trained filter $\Hdeblurhat$ to derive the deblurred norm map, $\hat{\C} = \Hdeblurhat * \S$, and compare it with the ground truth, the confocal norm map $\C$.
The histograms of the correlation values between $\C$ and $\hat{\C}$ in the $x$- and $y$-directions are shown using the ``deblurred'' legend of Figs.~\ref{fig:nm_corr_dist}(a) and (b), respectively.
Due to deblurring, the averaged correlations increased from $0.357$ to $0.442$ for the $x$-component and from $0.301$ to $0.396$ for the $y$-component. Their sample standard deviations also both decreased to $0.08$.
The increased correlations and decreased standard deviations after deblurring indicate that 
blurring is attributed to the lower quality of scanner-estimated norm maps.
It is also noted that, in light of the non-negligible but limited improvement of the correlation due to deblurring, more investigations are needed to reveal other factors limiting the accuracy of the scanner norm maps.
In the practical authentication system in Section \ref{sec:practical}, we do not apply deblurring, due to the limited improvement of correlation.

\subsubsection{Shape of Blurring Filter}
It is also interesting to estimate the blurring filter in order to directly reveal the characteristics of blurring.
First, we use a nonparametric approach to determine the shape of the blurring filter, which can avoid bias due to imposing a parametric model that may potentially cause mismatch.
We estimated a $7$-by-$7$ LSI filter $\Hblur$ such that $||\S - \Hblur * \C||_F^2$ was minimized. 
Since the coefficients in the blurring filter should all be non-negative, we estimated the blurring filter $\Hblur$ using non-negative least-squares.
Because the blurring filter has a lowpass nature and is an inverse filter of the deblurring filter, even a filter smaller than $7$-by-$7$ should be sufficient for adequately capturing the blurring effect.

\begin{figure}[!t]
\centering
  \vspace{-3mm}
  \hspace{-2mm}
  \subfloat[]{\includegraphics[width=0.48\linewidth]{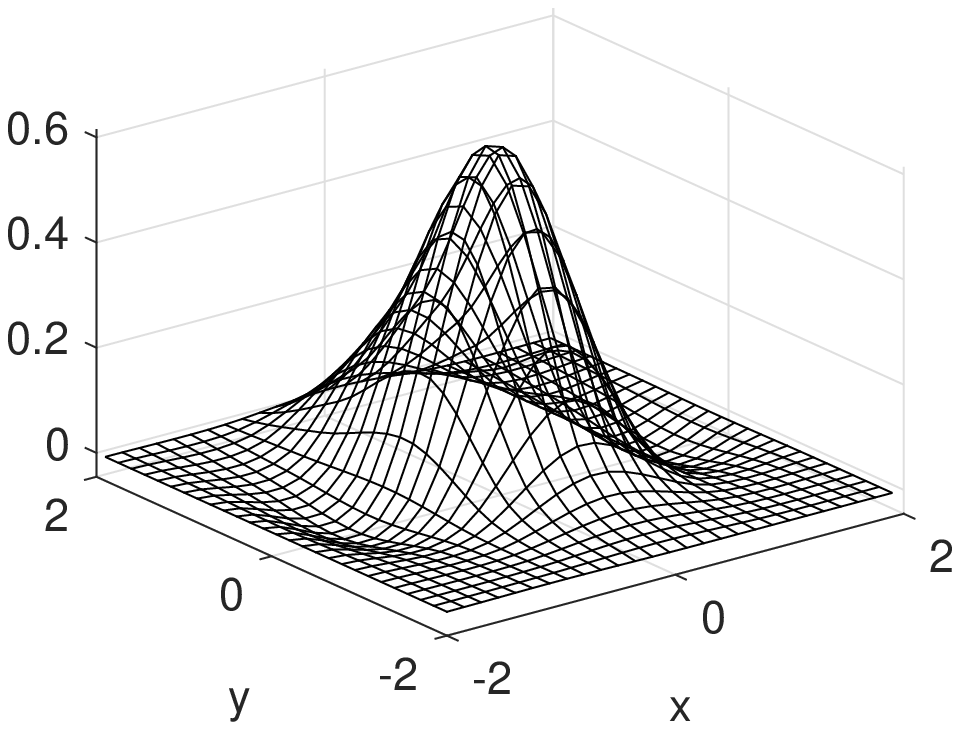}}
  \hspace{3mm}
  \subfloat[]{\includegraphics[width=0.48\linewidth]{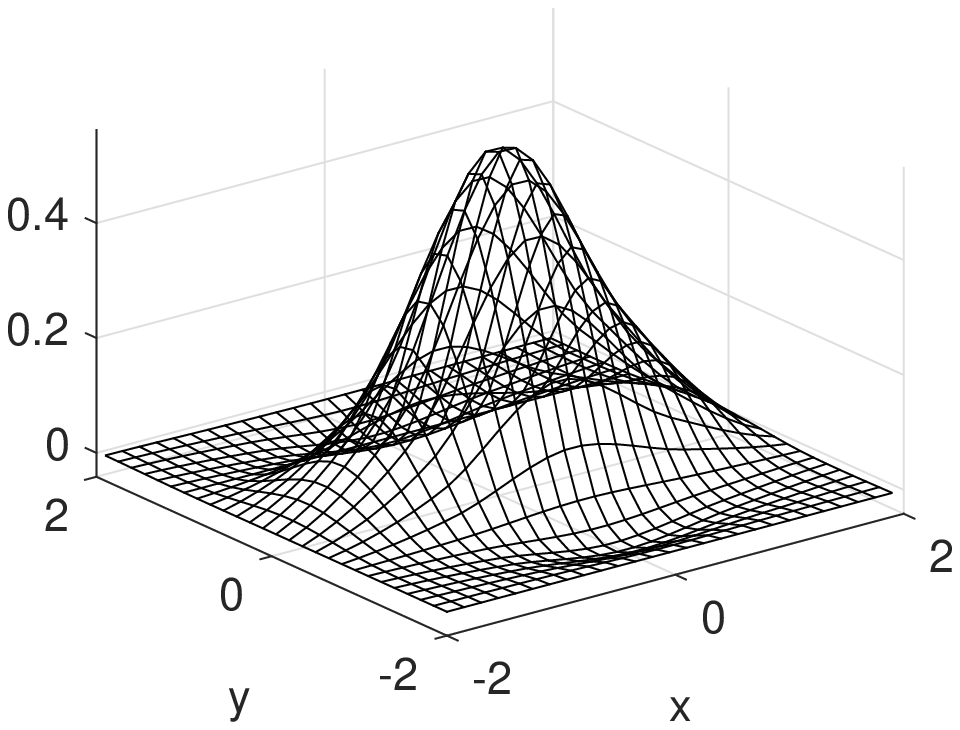}}
  
  \vspace{-0mm}
  \hspace{-2mm}
  \subfloat[]{\includegraphics[width=0.48\linewidth]{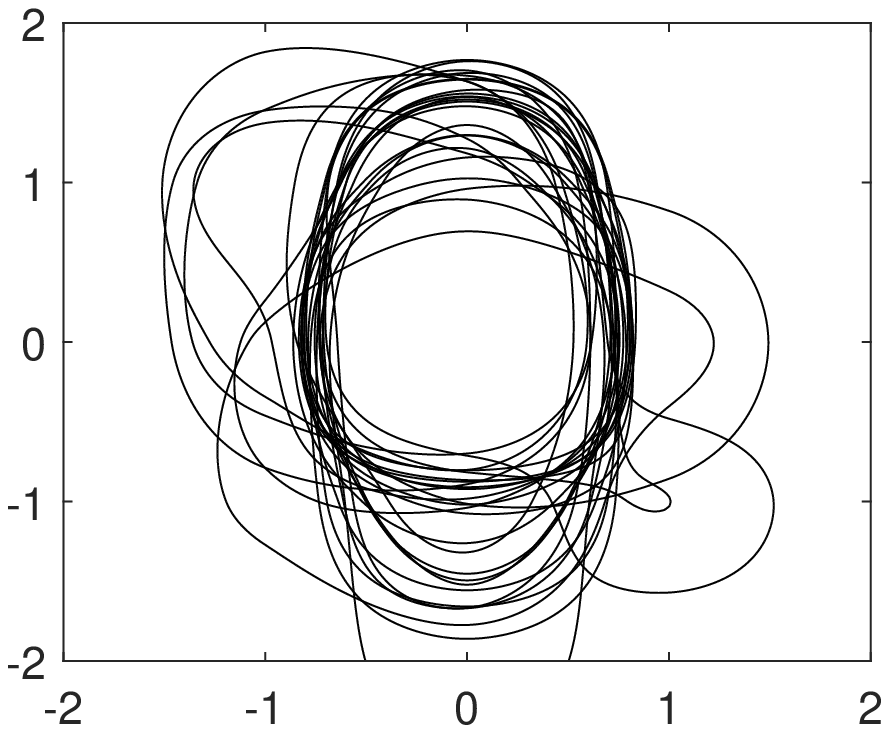}}
  \hspace{3mm}
  \subfloat[]{\includegraphics[width=0.48\linewidth]{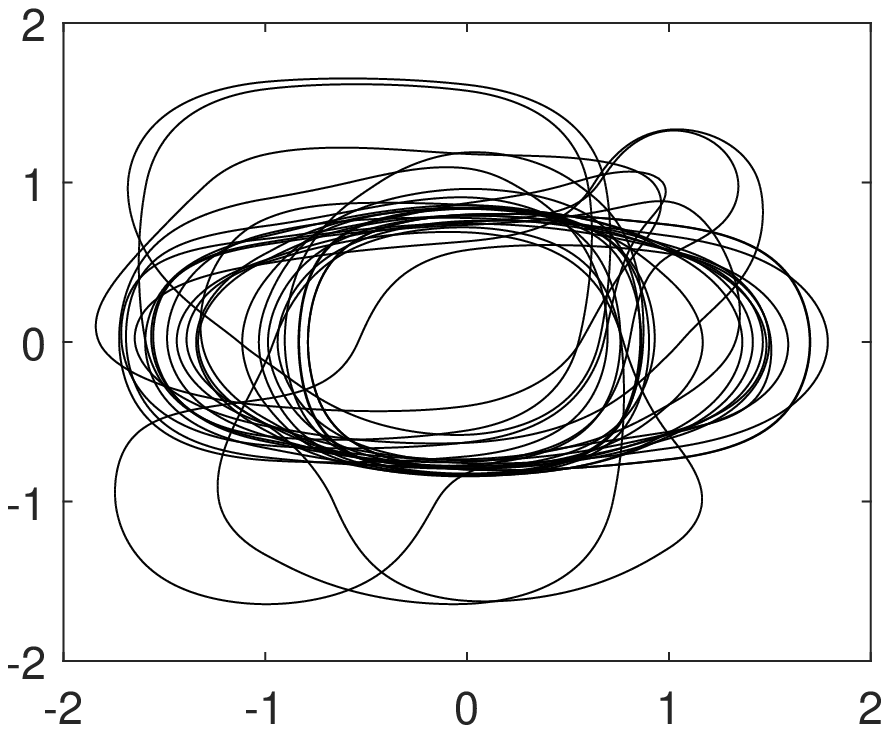}}
  \vspace{-2mm}
  \caption{Typical 3D mesh for the blurring filter for the (a) $x$- or (b) $y$-component of the norm map of a paper patch. 
  We also overlay the contour graphs (one contour per contour graph) for all nine paper patches to illustrate the shape of blurring filters for the (c) $x$- or (d) $y$-component of norm maps. 
  The blurring filters for the $x$-component of norm maps have larger variance in the $y$-direction and the blurring filters for the $y$-component of norm maps have larger variance in the $x$-direction.
  }
  \vspace{-3mm}
  \label{fig:filter_contuors}
\end{figure}

After obtaining an estimate of the blurring filter defined on a $7$-by-$7$ grid, we interpolated the filter spatially and drew the 3D meshes and contours/level curves to visualize its shape.
Figs.~\ref{fig:filter_contuors}(a) and (b) depict two typical 3D meshes for blurring filters derived from the $x$- and $y$- components of the norm map of one paper patch, respectively.
Figs.~\ref{fig:filter_contuors}(c) and (d) show one contour per filter for all paper patches used in our experiments.
The shapes of the contours reveal that the blurring filters for the $x$-component of the norm maps have larger spread in the $y$-direction and the blurring filters for the $y$-component of the norm maps have larger spread in the $x$-direction.
The shapes of the contours are similar, so different scanners have similar blurring effects.

\begin{figure}[!t]
\centering
  \vspace{-0mm}
  \hspace{-2mm}
  \subfloat[]{\includegraphics[width=0.48\linewidth]{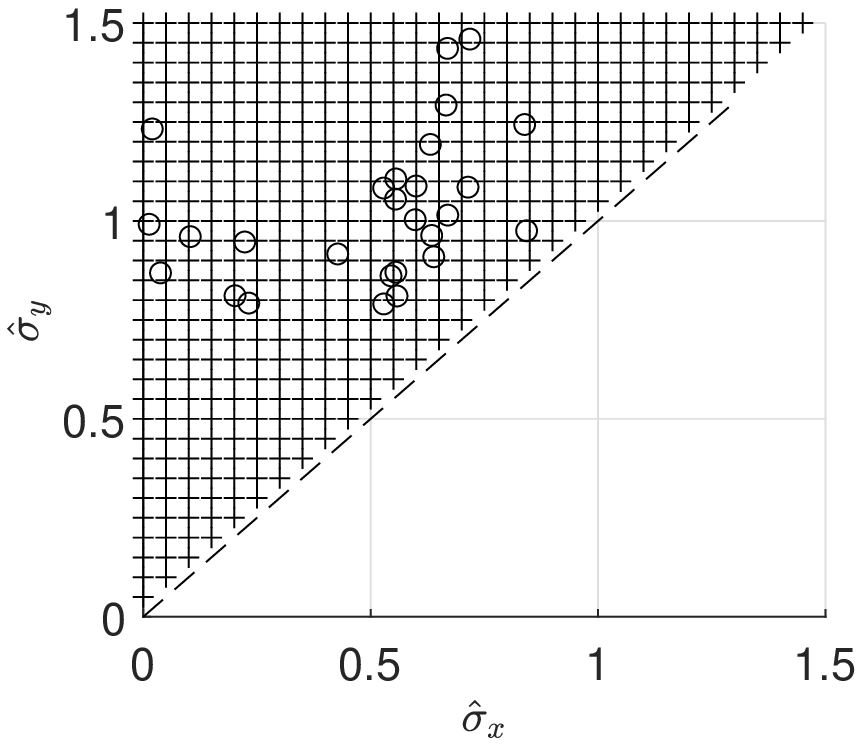}}
  \hspace{2mm}
  \subfloat[]{\includegraphics[width=0.48\linewidth]{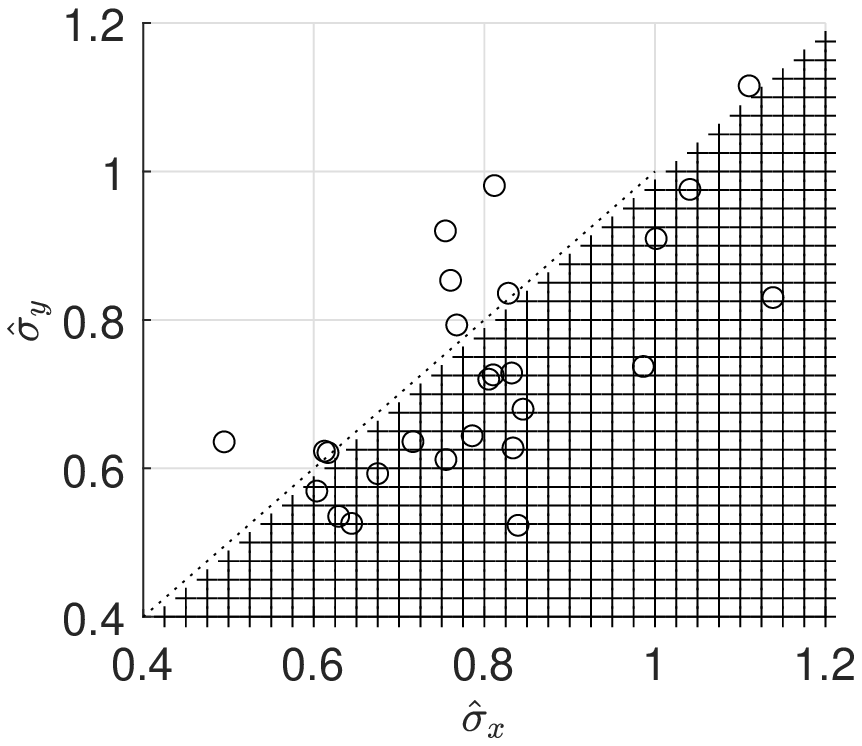}}
  \caption{Scatter plot of $(\hat{\sigma}_x, \hat{\sigma}_y)$ for the blurring filters of the (a) $x$- or (b) $y$-component of the norm maps for all nine paper patches.
  In the $x$-component of norm map the variance in the $x$-direction is smaller, and in the $y$-component of norm map the variance in the $y$-direction is in general smaller, as illustrated through the shaded regions. }
  \vspace{-4mm}
  \label{fig:G_filter_sigmas}
\end{figure}

Since the blurring filters are close to bell-shaped, we further obtain a quantitative description of the spread for the blurring filters using parametric Gaussian filters.
Let us assume a blurring filter that is generated by discretizing and normalizing a separable bivariate Gaussian function on a $7$-by-$7$ grid.
The bivariate Gaussian is parameterized by $\mu_x, \mu_y, \sigma_x, \sigma_y$,
where $(\mu_x, \mu_y)$ describes the location of the filter, 
and $\sigma_x$ and $\sigma_y$ are the standard deviations of the Gaussian filter in the $x$ and $y$ directions.
We assume the Gaussian to be separable based on the fact that blurring in the $x$ and $y$ directions have different causes due to the geometry of the flatbed scanner, and the observations from Fig.~\ref{fig:filter_contuors} that nonparametrically estimated filters' contours are oriented horizontally or vertically. 
We estimate $\Hblur^{\text{Gaussian}} = \G(\mu_x, \mu_y, \sigma_x, \sigma_y)$ by solving the following minimization problem:
\begin{equation}
\min_{\mu_x, \mu_y, \sigma_x, \sigma_y} ||\S - \G(\mu_x, \mu_y, \sigma_x, \sigma_y) * \C||_F^2 .
\end{equation} 
Since this problem is nonconvex, we numerically solve it with the following starting point configurations by taking into consideration the nonparametric results summarized in Fig.~\ref{fig:filter_contuors}:  $\sigma_x=\sigma_y=1$, and $\mu_x$, $\mu_y$ are uniformly randomly drawn from $-0.5$ to $0.5$.
The estimated standard deviations in the $x$- and $y$-components of the norm maps for different paper patches are shown in Fig.~\ref{fig:G_filter_sigmas}, which are consistent with the results in Fig.~\ref{fig:filter_contuors}.

The results of parametric Gaussian filters confirmed the following patterns obtained from the nonparametric least-squares method: i) the variance in the $x$-direction is smaller for the $x$-component of the norm map, and ii) the variance in the $y$-direction is smaller for the $y$-component of the norm map.
Note that a smaller variance indicates a weaker blurring effect.
This phenomenon may be due to the unique optical setup of Contact Image Sensor (CIS) flatbed scanners \cite{cisscanner}. The three flatbed scanners used in the experiments of this paper are all CIS flatbed scanners. CIS scanners are equipped with a gradient-index lens array whose focal length is only around $0.1$ mm \cite{gorocs2014biomedical}. According to Fig.~1(b) of \cite{liu2018enhanced}, the range of height of paper surface is more than $0.1$ mm from the measurements using a confocal microscope. This will introduce out-of-focus blur when the images of paper patches are acquired by CIS scanners. Since a strip of paper is completely lit along the linear light direction and has limited light spread along the scanning direction, more blurring will be preserved along the direction of the linear light as the scanned image is created by stitching the scanned lines.

\section{Heightmap as a Discriminative Feature} \label{sec:3D}
Although the norm map has been shown to be a powerful discriminative feature \cite{clarkson-09, wong2017}, 
when it is used in a practical authentication system it is desirable to further increase the discriminative power to ensure a better performance. 
Previous work in \cite{liu2018enhanced} used the estimated norm map to reconstruct the heightmaps (3D surfaces) and discovered that high-frequency subbands of
reconstructed heightmap are more powerful than the norm map in describing the uniqueness of a physical surface.
The result in \cite{liu2018enhanced} was demonstrated for mobile cameras and, in this section, we investigate whether a similar conclusion can be drawn for flatbed scanners.

\subsection{$Z$-Component Estimation From Norm Map} 
\begin{figure}[!t]
\centering
  \vspace{-3mm}
  \hspace{-0mm}
  \subfloat[]{\includegraphics[width=0.48\linewidth]{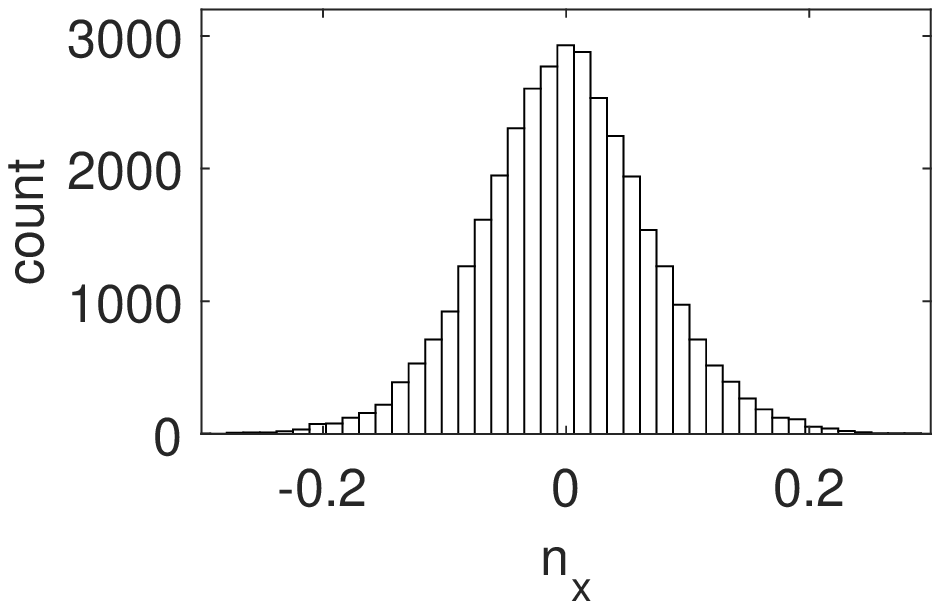}}
  \subfloat[]{\includegraphics[width=0.48\linewidth]{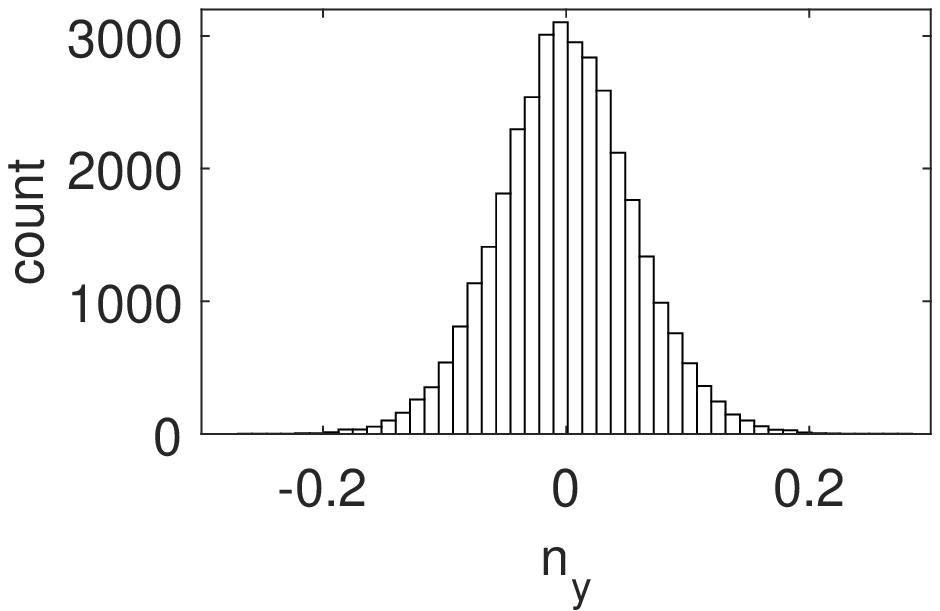}}
  \vspace{-0mm}
  \hspace{-0mm}
  \subfloat[]{\includegraphics[width=0.48\linewidth]{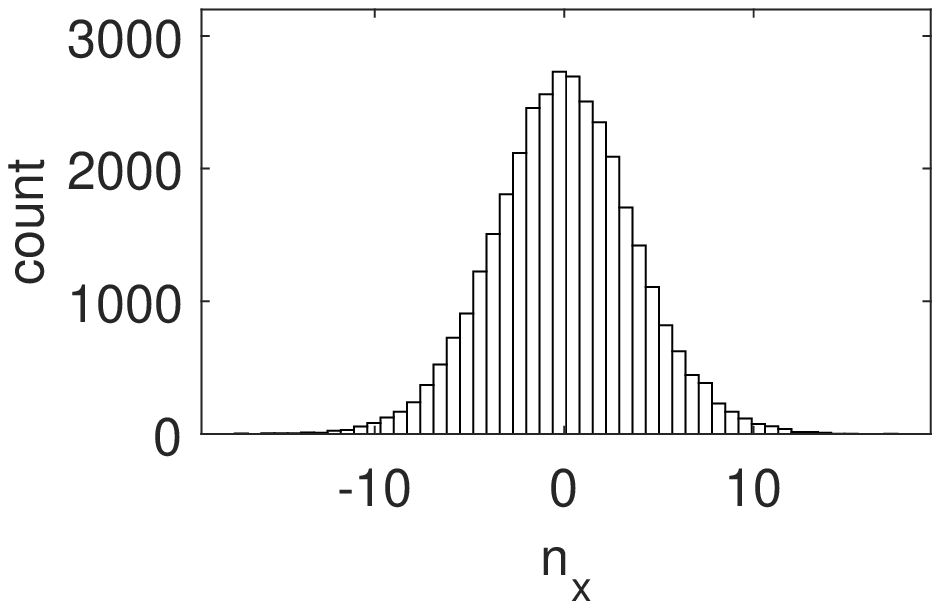}}
  \subfloat[]{\includegraphics[width=0.48\linewidth]{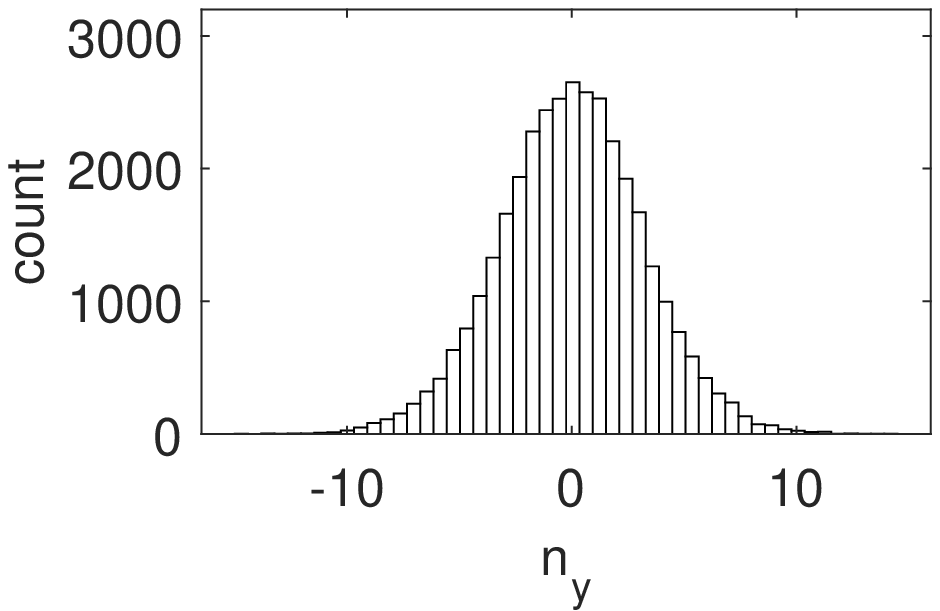}}
  \caption{Histograms for the (a) $x$- and (b) $y$-components of norm map from confocal microscope. 
  Histograms for the (c) $x$- and (d) $y$-components of norm map from scanner. Note that the components calculated from scanners are off by an unknown scaling factor.
  The distributions are Gaussian-like and roughly centered around zero.
	}
  \vspace{-3mm}
  \label{fig:nx_ny_dist}
\end{figure}

In this subsection, we propose an estimator for the $z$-component of the normal vector field based on a known norm map for surface reconstruction. 
Surface reconstruction in general requires a normal vector field containing at each location a 3-D description about the orientation \cite{book, reconstruct}.
However, using images acquired by scanners and the estimation technique presented in Section~\ref{sec:deviations}, only the norm map, i.e., the scaled versions of the $x$- and $y$-components of the normal vector field, $(n_x^{(s)}, n_y^{(s)})$, are available.
The authors of \cite{liu2018enhanced} proposed a distribution matching approach to estimate scalars $\alpha_x$ and $\alpha_y$ that correctly normalizes the norm map so that the $z$-component can be calculated using $\hat{n}_z = \left[ 1- (n_x^{(s)}/\hat{\alpha}_x)^2 - (n_y^{(s)}/\hat{\alpha}_y)^2 \right]^{1/2}$, where the quantities with hats are the corresponding estimated values. The distribution matching approach finds the best $\hat{\alpha}_x$ and $\hat{\alpha}_y$ such that the standard deviations of $n_x^{(s)}/\hat{\alpha}_x$ and $n_y^{(s)}/\hat{\alpha}_y$ will match those of the confocal. However, details for obtaining $\hat{\alpha}_x$ and $\hat{\alpha}_y$ were not given. Below, we justify the approach proposed in \cite{liu2018enhanced} and propose a least-squares formula for estimating a shared scalar $\alpha$ for both directions.
We first examine the real data to support subsequent model design. We show histograms for the $x$-, $y$-, and $z$-components of the normal vector field in Figs.~\ref{fig:nx_ny_dist}(a), \ref{fig:nx_ny_dist}(b), and \ref{fig:nz_dist}(a), respectively.
From the histograms, we can see that normal vectors are on average pointing straight up, due to large $n_z$ and are without obvious bias in both the $x$- and $y$-directions.
The distributions are Gaussian-like and centered around zero. 
We also plot the histograms for $x$- and $y$-components of the norm map that are scaled. 
We observe that they are similarly distributed to those from the confocal but are scaled, centered around $0$.
The above observation on the real data implies that a scaling relation is enough to connect the norm map to the first two components of the normal vector field, namely $n_x^{(c)}$ and $n_y^{(c)}$. 
Since the $x$- and $y$-components of the norm map are obtained by the same scanning process with the only difference being scanning directions, a shared multiplicative scalar should be used for both dimensions, namely, 
\begin{subequations}
\begin{align}
(n_x^{(s)}, n_y^{(s)}) &\approx \alpha \cdot (n_x^{(c)}, n_y^{(c)}), \label{eq:norm_vec_scaling_raw}\\
(\sigma_x^{(s)}, \sigma_y^{(s)}) &\approx \alpha \cdot (\sigma_x^{(c)}, \sigma_y^{(c)}), \label{eq:norm_vec_scaling_std}
\end{align}
\end{subequations}
where \eqref{eq:norm_vec_scaling_std} was obtained by considering $\alpha$ as a constant and other components in \eqref{eq:norm_vec_scaling_raw} as random variables, and by applying the variance operation to both sides of \eqref{eq:norm_vec_scaling_raw}.
Estimating $\alpha$ using least-squares from \eqref{eq:norm_vec_scaling_std}, we obtain
\begin{equation}
\hat{\alpha} = ({\sigma_x^{(s)} \sigma_x^{(c)} + \sigma_y^{(s)} \sigma_y^{(c)}}) \Big/ ({{\sigma_x^{(c)}}^2 + {\sigma_y^{(c)}}^2}),
\label{eq:alpha_eq}
\end{equation}
which blends in the scaling effect in both directions. This formula allows the calculation of a scalar for a scanner norm map by using merely two summary statistics of the paper surface, $\sigma_x^{(c)}$ and $\sigma_y^{(c)}$, that are determined by the physical characteristics of papers and are stable numbers for papers of the same type \cite{beland2000effect}.

\begin{figure}[!t]
\centering
  \vspace{-0mm}

  \hspace{-0mm}
  \includegraphics[width=\linewidth]{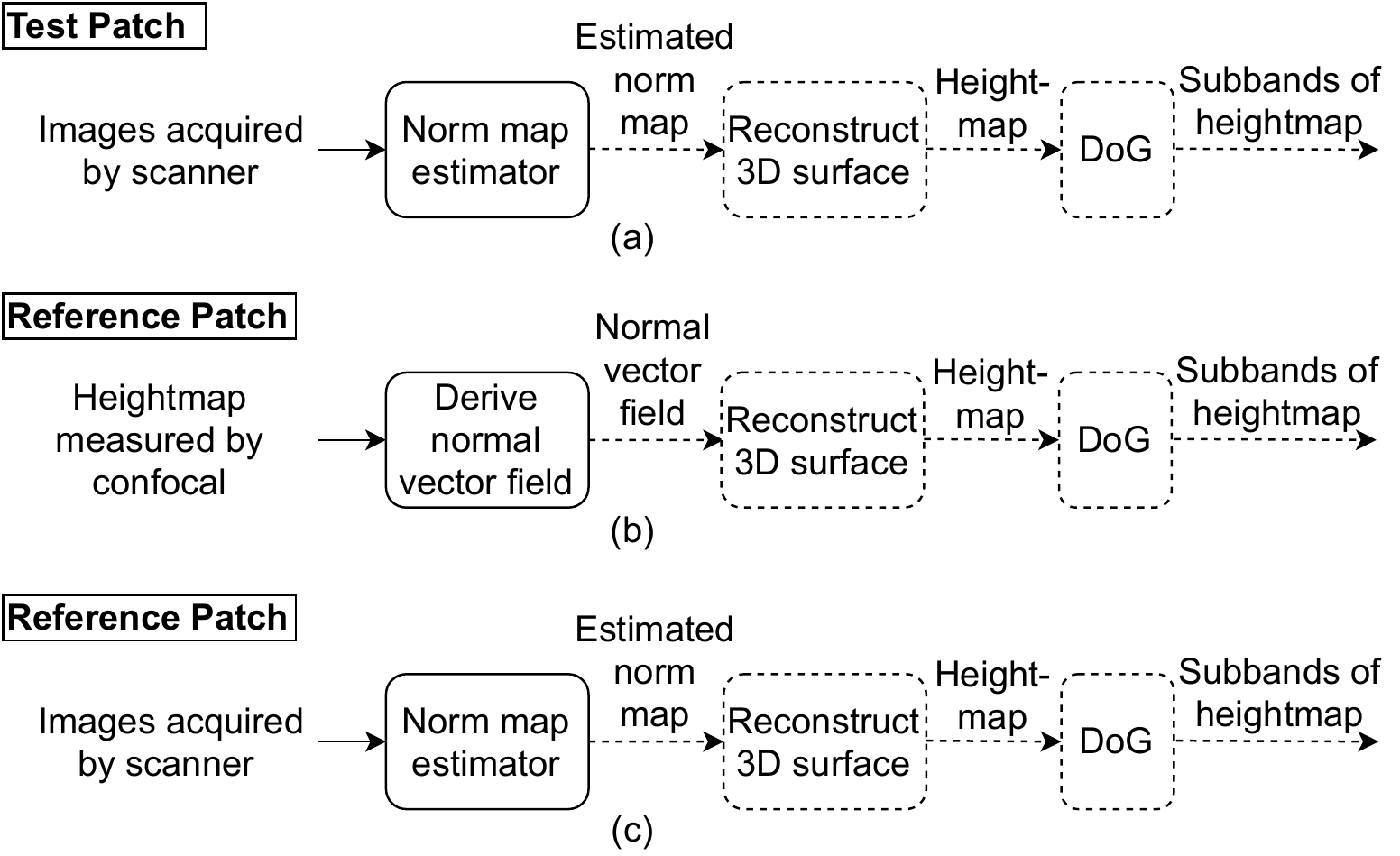}
  \caption{(a) Block diagram for obtaining features from a test patch using images acquired by a flatbed scanner. Block diagrams for obtaining features from a reference patch using (b) measurements from a confocal microscope, or (c) images acquired by a flatbed scanner. The norm map, the heightmap, or the subbands can be used as discriminative features. The blocks/processes with dashed boundaries should be ignored when their inputs are used as features.}
  \vspace{-4mm}
  \label{fig:diagram}
\end{figure}

\subsection{Heightmap and Subbands as Discriminative Features} \label{subsec:heightmap_subbands}
In \cite{liu2018enhanced}, the authors have shown experimentally for mobile camera-acquired images that the high frequency subbands have been proven to be powerful discriminative features for authentication.
In this work, we validate the method of \cite{liu2018enhanced} using flatbed scanner-acquired images.
We follow the procedure in \cite{liu2018enhanced} to reconstruct 3D heightmaps of paper patches and derive the subbands of the reconstructed heightmaps for authentication. 
The reference data are from a confocal microscope. Each paper patch was scanned once by the confocal microscope. The test data are obtained from scanners. Each paper patch was scanned by one scanner three times, and there were three different scanners used. Thus, each paper patch has one ground-truth heightmap from the confocal microscope and nine reconstructed heightmaps from scanners.

Here, we summarize the benefit of using detrended heightmaps, and more details are given in Section~A of the supplementary document. The correlation value using reconstructed heightmaps improved to $0.358$ from $0.357$, or $0.301$, when using the norm map as the discriminative feature, as shown in Table~\ref{tab:corr_summary}. When using the detrended heightmap as the discriminative feature, the correlation value further improved to $0.499$. This result is consistent with that reported in \cite{liu2018enhanced}, in which a mobile camera was used as the acquisitions device. Hence, the detrended heightmap is a more powerful discriminative feature than the norm map.

\begin{figure}[!t]
\centering
  \vspace{-3mm}
	\subfloat[]{\includegraphics[width=0.48\linewidth]{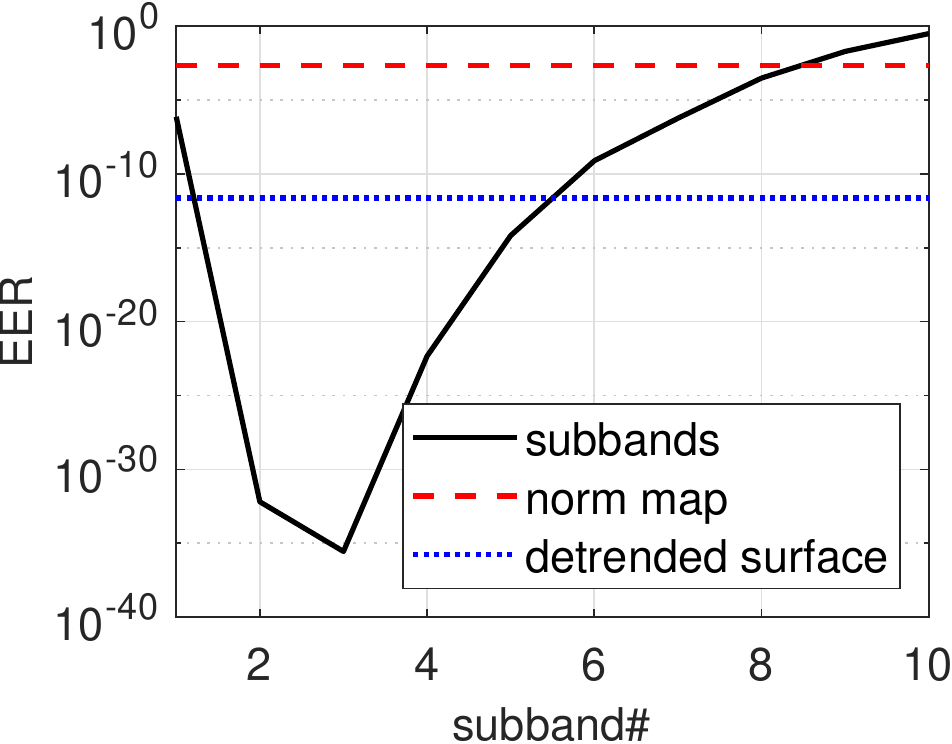}}
	\hspace{2mm}
	\subfloat[]{\includegraphics[width=0.48\linewidth]{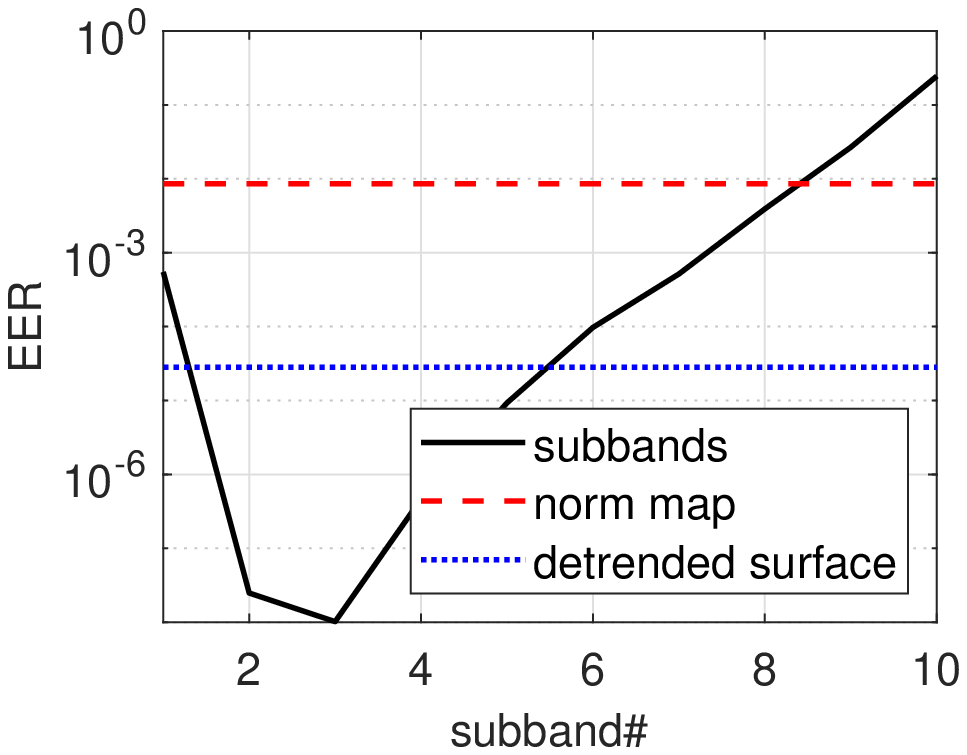}}
  \caption{EER calculated for every subband when correlation values are believed to follow (a) Gaussian or (b) Laplace distributions. The reference data are obtained by a confocal microscope and the test data are acquired with flatbed scanners. The third-highest spatial-frequency subband is the most powerful in describing the uniqueness of physical surfaces. Horizontal lines correspond to the performance when the norm map or detrended surface/heightmap is used as the discriminative feature. }
  \vspace{-3mm}
  \label{fig:eer}
\end{figure}

We also summarize the benefits of using high spatial-frequency subbands of heightmaps, and more details are given in Section~B of the supplementary document.
We decomposed the reconstructed heightmap into ten spatial subbands corresponding to a DoG representation, as reviewed in Section \ref{subsec:dog}.
Using the third-highest spatial-frequency subband instead of the detrended heightmap, the correlation value improved from $0.499$ to $0.714$.
The estimated EER as a function of subband index is shown in Fig.~\ref{fig:eer}, where it may be seen that a smaller subband index corresponds to a higher spatial frequency. When using the third-highest spatial-frequency subband, the EER achieved $10^{-36}$ or $10^{-8}$ under the Gaussian or Laplacian tail extrapolation assumption, respectively, either of which constitute a large improvement over $10^{-11}$ or $10^{-4.5}$ when using a detrended heightmap. The high spatial-frequency subbands are more powerful discriminative features than detrended heightmaps when using flatbed scanners for paper surface-based authentication.

\subsection{Practical Authentication System} \label{sec:practical}
In this subsection, we examine a practical authentication system that uses flatbed scanners to acquire both test and reference data.
Using every subband of the heightmap as the discriminative feature, we compare to the traditional feature, i.e., to the norm map, and measure the authentication performance in EER.
The diagrams for generating the subbands in the authentication system for test and reference patches are shown in Figs.~\ref{fig:diagram}(a) and (c), respectively.
The reference and test patches are both images acquired by scanners.

In the practical authentication system, we use scanners instead of the confocal microscope to capture the reference data because scanners are easier to automate and more affordable for practical deployment. 
Each paper patch was scanned three times by each of the three scanners. We obtained nine norm maps using scanners for each paper patch. For the matched case, we chose two norm maps from the nine norm maps each time as a test-reference pair, forming a total of $36$ pairs for each paper patch. Given the nine physical pieces of paper patches, this leads to a total of $36 \times 9 = 324$ data points of correlation values for statistical analysis. For the unmatched case, each paper patch pair gives $9 \times 9 =81$ data points, and there are $\binom92 = 36$ paper patch pairs. Theoretically, there are in total $81 \times 36 = 2916$ data points, if using all paper patches. To mimic a practical scenario, we randomly chose one paper patch from the rest of the paper patches from which to obtain the reference data, leading to a random subset of $729$ data points for the unmatched case.

\begin{figure}[!t]
\centering
  \vspace{-3mm}
	\subfloat[]{\includegraphics[width=0.48\linewidth]{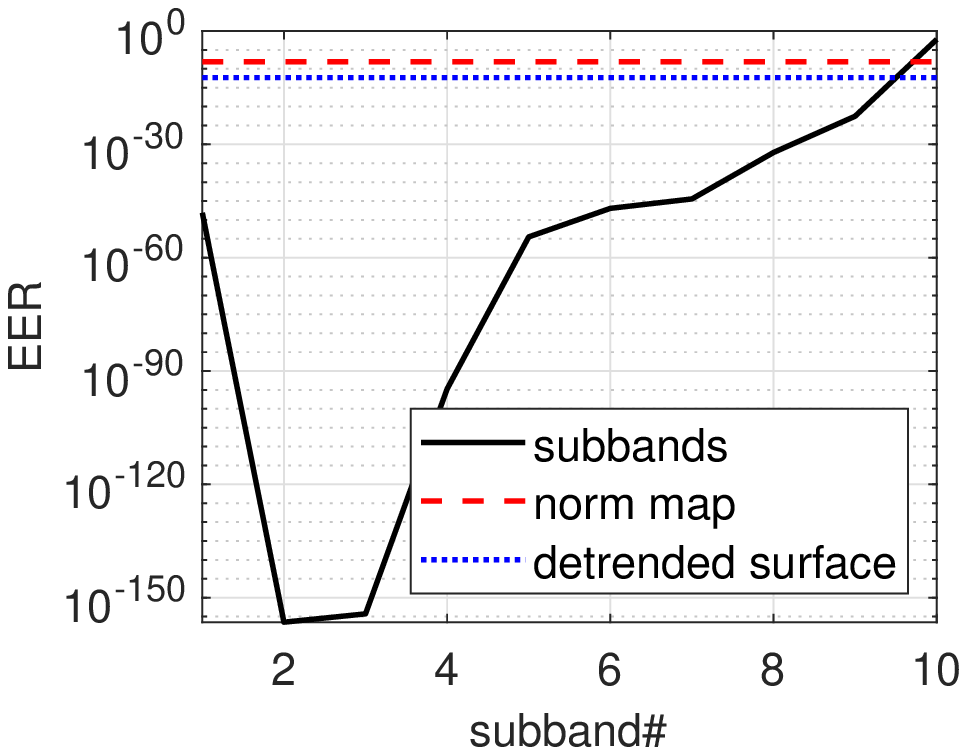}}
	\hspace{2mm}
	\subfloat[]{\includegraphics[width=0.48\linewidth]{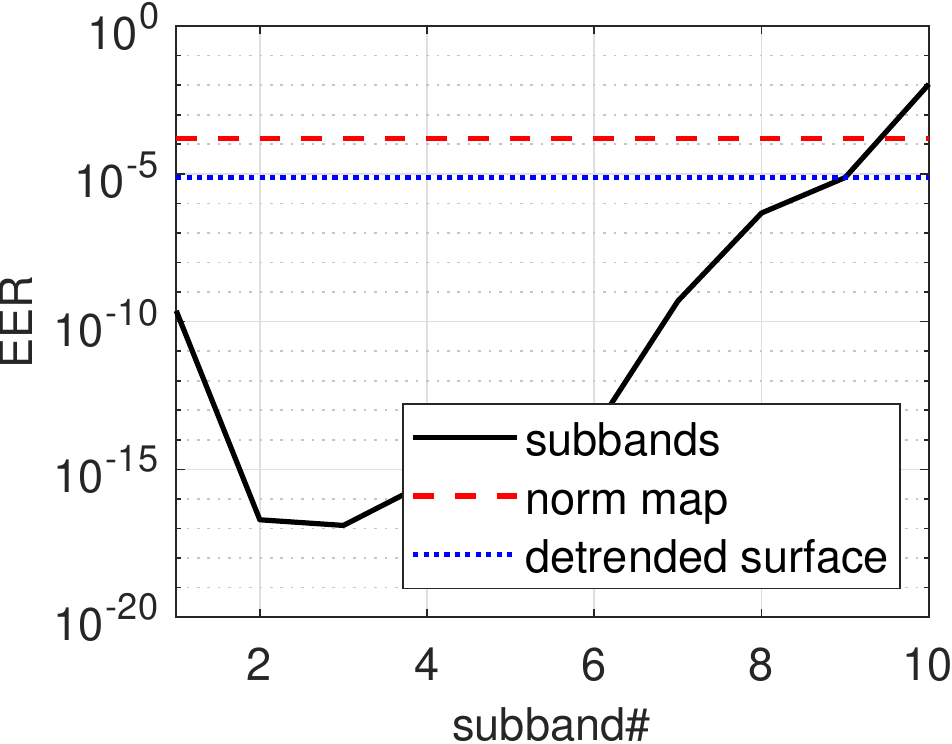}}
  \caption{EER calculated for every subband when correlation values are believed to follow (a) Gaussian or (b) Laplace distributions.
  The second-highest spatial-frequency subband has the most powerful authentication capability in a practical setup whereby scanners are used to acquire reference data.
  Horizontal lines correspond to performance when the norm map or detrended surface/heightmap is used as the discriminative feature.}
 \vspace{-3mm}
  \label{fig:eer_scanners}
\end{figure}

We reconstructed 3D surfaces from the norm maps and obtained the subbands of the heightmap as the discriminative features. We calculated the correlation values of subbands between the test and reference data.
We calculated the EER for every subband and plotted the results in Fig.~\ref{fig:eer_scanners}.
When correlation values are believed to follow Gaussian or Laplace distributions, the EER are about $10^{-157}$ and $10^{-17}$, respectively, at the second-highest spatial-frequency subband.
We also compared the performance of subbands of the heightmap to that of the norm map and detrended heightmap, as shown by horizontal lines in Fig.~\ref{fig:eer_scanners}, from which we found that EERs using the norm maps and the detrended surfaces are much larger than using the second-highest spatial-frequency subband.
Hence, in the practical system that uses a scanner to acquire reference data, the authentication performance of the second-highest spatial-frequency subband is much better than that of the norm map or detrended surfaces, in terms of EER.
In Table \ref{tab:practical_system_summary}, we summarized the authentication performance of the practical authentication system in this work. For comparison, we also reproduced the results in \cite{liu2018enhanced} where mobile cameras, instead of scanners, were used to obtain the test data. We compared the best EER of the subbands when assuming the correlation values are Gaussian and Laplacian distributed. The performance of the practical authentication system in this work using scanners to obtain test data is much better than using a mobile camera in terms of EER.
\begin{table}[!t]
  \caption{Comparison of Performance of Practical Authentication System When Test Data is Obtained from Mobile Camera or Scanner\vspace{-2mm}}
  \label{tab:practical_system_summary}
  \centering
  \scalebox{1.0}{
  \begin{tabular}{lclcl}
\hline \hline
 Test  & Reference   & Feature &  EER (Gaussian \\ 
 device &    device  &         &   and Laplacian)  \\ \hline
 mobile camera & scanner & norm map &   $10^{-5}$ and $10^{-3}$\cite{liu2018enhanced} \\
 mobile camera  & scanner & subband &   $10^{-8}$ and $10^{-3}$\cite{liu2018enhanced} \\
 scanner & scanner & norm map &   $10^{-9}$ and $10^{-4}$\\ 
 scanner & scanner & subband &  $10^{-157}$ and $10^{-17}$ \\ 
\hline
\hline
  \end{tabular}
	}
	\label{tab:parctical_system}
  \vspace{-1mm}
\end{table}

\section{Size of Paper Patch, Digitization Resolution, and Perturbation of Alignment} \label{sec:discussion}

\subsection{How Large Should the Size of the Paper Patch Be?} \label{subsec:paper_size}
\begin{figure}[!t]
\centering
  \vspace{-3mm}
	\subfloat[]{\includegraphics[width=0.48\linewidth]{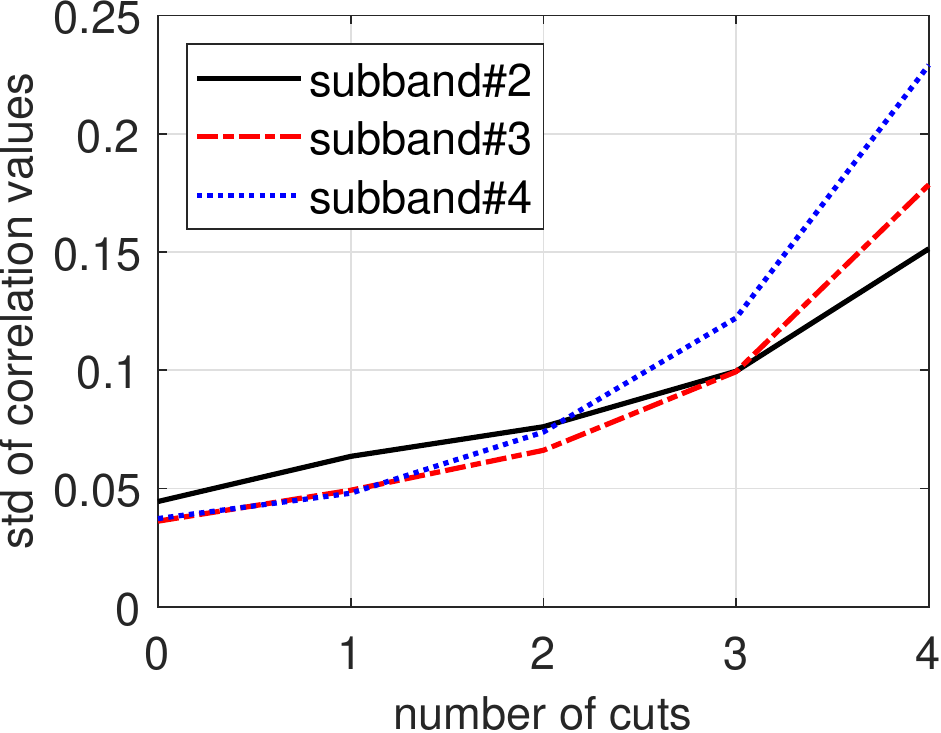}}
	\hspace{2mm}
	\subfloat[]{\includegraphics[width=0.48\linewidth]{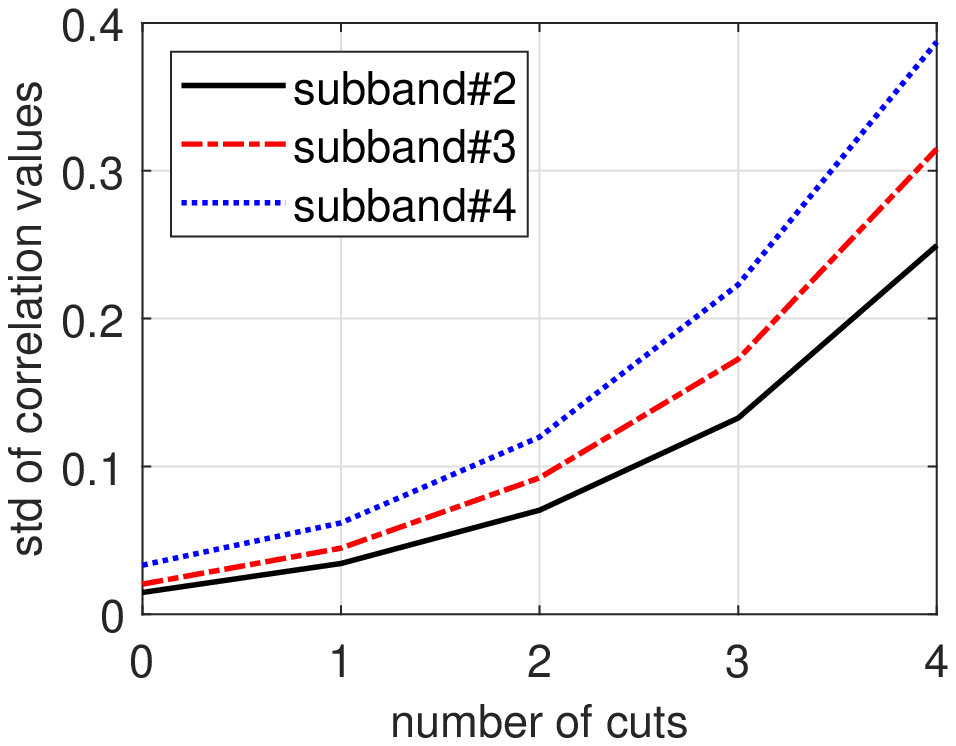}}
  \caption{Sample standard deviations of the correlation values when cutting the paper patch into blocks under (a) matched and (b) unmatched cases.
 The standard deviations of the correlation values in spatial-frequency subbands \#2--\#4 increase exponentially when cutting paper patches into small blocks.}
\vspace{-3mm}
  \label{fig:std_rho}
\end{figure}

\begin{figure}[!t]
\centering
  \vspace{-0mm}
	\subfloat[]{\includegraphics[width=0.48\linewidth]{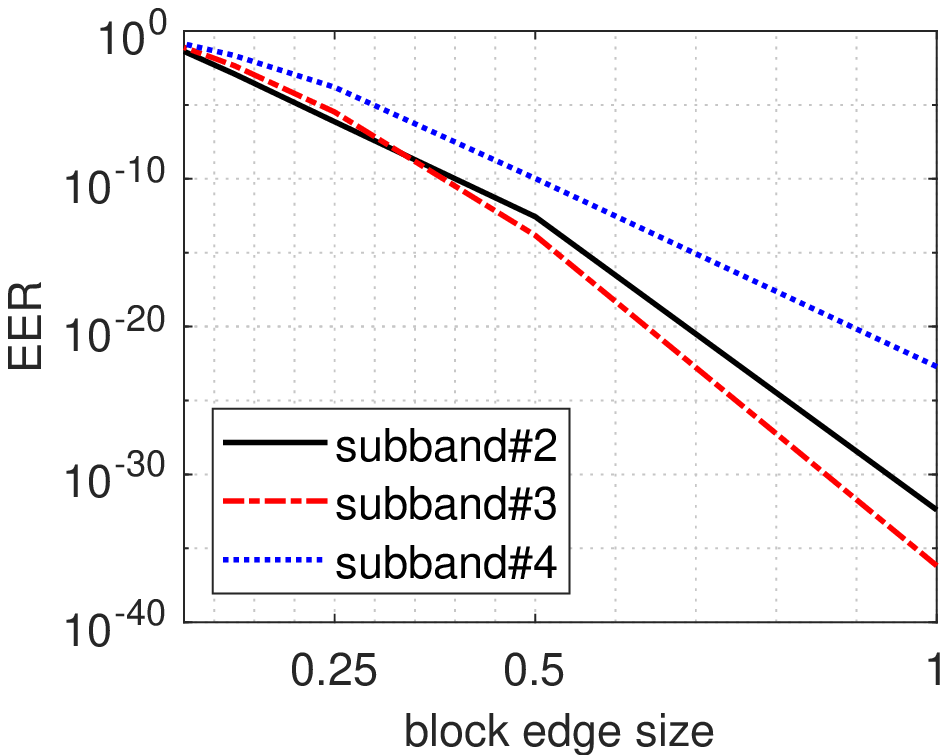}}
	\hspace{2mm}
	\subfloat[]{\includegraphics[width=0.48\linewidth]{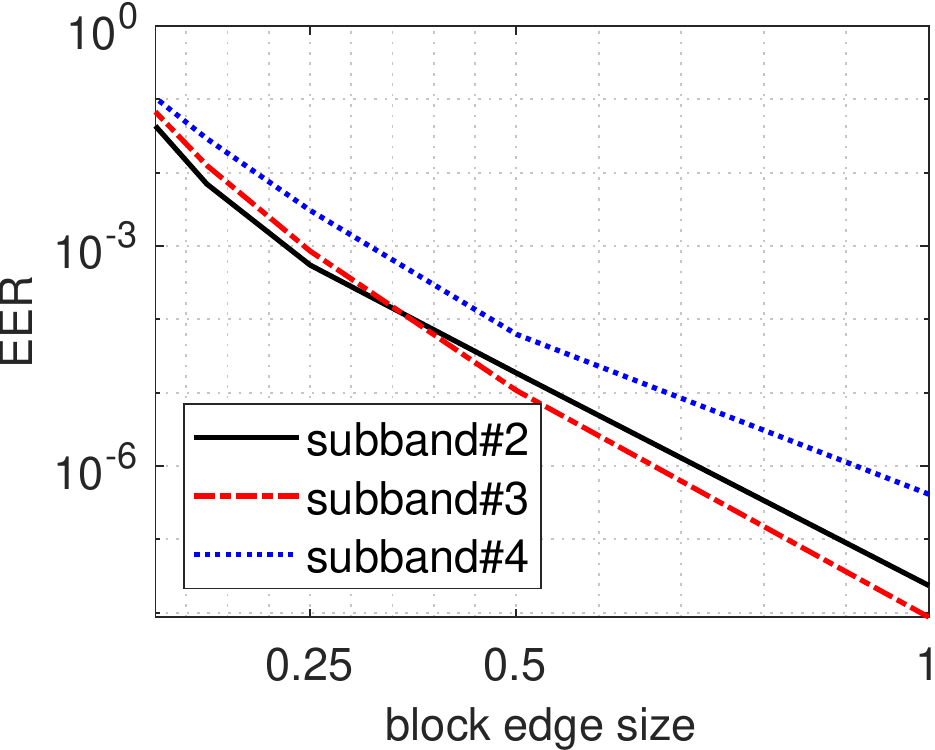}}
  \caption{After cutting paper patches into blocks, EERs against the block edge length when assuming (a) Gaussian and (b) Laplace distributions. Size of $1$ corresponds to the edge length of the original patch.
 The EERs decrease when the block edge length increases.}
  \label{fig:eer_block}
\end{figure}

Throughout the experiments of this work, the size of the paper patch was fixed to be $\frac{2}{3}$-by-$\frac{2}{3}$ inch$^2$ and discretized to $200$-by-$200$ pixels.
A natural research question pertaining to a practical deployment is: How does the size of the paper patch affect the authentication performance?
To investigate this question, we successively cut one heightmap into four heightmaps, empirically calculated the EER using the smaller heightmaps after each cut, and examined how the EER changes as the number of cuts increases.
More specifically, we regarded the heightmap's center $160$-by-$160$ pixels as the root patch that had not been cut.
After the first cut, the resulting heightmaps were of the size $80$-by-$80$ pixels.
At each cut level, we calculated the correlation values against confocal references.
We observed that, after each cut, the means of correlation values were almost unchanged, whereas the standard deviation would increase by a factor of $\sim 2$ times for unmatched cases and $\sim 1.5$ times for matched cases.
We plotted the sample standard deviations of the correlation values as a function of the number of cuts in Fig.~\ref{fig:std_rho}.
We further calculated EERs at each cutting level and plotted EERs against the block edge size in Fig.~\ref{fig:eer_block}, in which a block edge size $= 1$ corresponds to using 160 pixels.
As expected, the authentication performance in EER improves as the block size increases. 

Below, we analytically show that the EER is exponentially decreasing in the size of the paper patch when correlation values are assumed to be Laplacian distributed.
Using the EER formula (1b) provided in supplementary document and the variance formula of a Laplace random variable, $\lambda = \sqrt{2}/\sigma$, the EER can be rewritten as $\EER = \frac{1}{2}\exp \big[ \frac{\sqrt{2}}{\sigma_0+\sigma_1}(\mu_0-\mu_1) \big]$.
After $n$ cuts, the ERR can be expressed as 
\begin{subequations}
\begin{align}
\EER(n) &= \frac{1}{2}\exp \Big[ \frac{\sqrt{2}}{2^n \sigma_0 + 1.5^n \sigma_1}(\mu_0-\mu_1) \Big] \label{eq:eer_n_1}\\
&\approx \frac{1}{2}\exp \big[ \sqrt{2} \cdot 2^{-n} (\mu_0-\mu_1)/\sigma_0  \big], 
\label{eq:eer_n_2}
\end{align}
\end{subequations}
where \eqref{eq:eer_n_1} incorporates the empirically observed exponential increase of the standard deviations in the previous paragraph, and \eqref{eq:eer_n_2} is approximately true for large $n$. 
Since $2^{-n}$ is proportional to the block edge size after $n$ cuts, $\log(\EER(n))$ is linearly decreasing in the block edge length, which is consistent with Fig.~\ref{fig:eer_block}(b).
When the edge length decreases from 160 pixels (or $0.53$ inches) to 80 pixels (or $0.27$ inches), the performance drops from around $10^{-9}$ to $10^{-5}$ in EER.
To conclude, a larger patch size will lead to better authentication performance, and given a certain paper type, experiments similar to the one demonstrated in this subsection may be conducted to determine the patch size needed to achieve a certain performance level.

\begin{figure}[!t]
\centering
  \vspace{-1mm}
	\includegraphics[width=\linewidth]{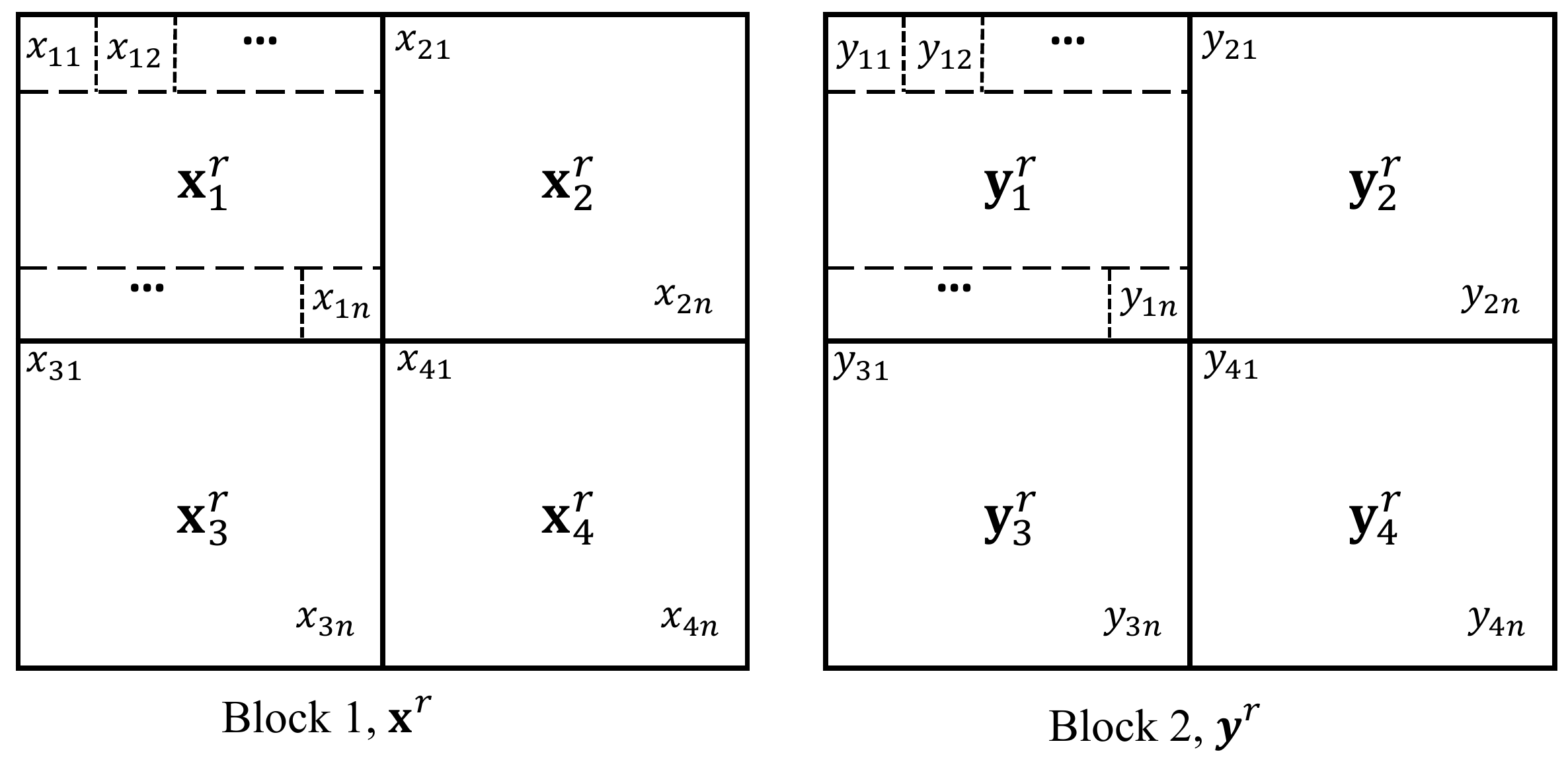}
  \caption{Sample correlation coefficients: $\rho$, between two blocks; $\rho_i$, between two collocated subblocks with index $i$. Detailed definitions are as follows: $\rho_i = \corr(\x_i^r, \y_i^r),\ i=1,\dots,4$, and $\rho = \corr(\x^r, \y^r)$, where the superscript ``$r$'' stands for the raw image data before the sample mean is removed. $\x_i^r$ and $\y_i^r$ are length-$n$ column vectors containing all pixel values of the respective subblocks. $\x^r$ and $\y^r$ are concatenated column vectors where $\x^r = (\x_1^r,\cdots,\x_4^r)$ and $\y^r = (\y_1^r,\cdots,\y_4^r)$.}
  \label{fig:rho_relation}
\end{figure}

Below, we justify the exponential increase of the standard deviation for correlation values as the number of cuts increases.
First, we claim the following finite-sample relation between the sample correlation coefficient of a block, $\rho$, and the sample correlation coefficients of its nonoverlapping, equal-sized subblocks, $\{\rho_i\}_{i=1}^4$, namely,
\begin{equation}
\rho \approx \frac{1}{4} \sum_{i=1}^4 \rho_i.
\label{eq:rho_approx_relation}
\end{equation}
The blocks and subblocks are illustrated in Fig.~\ref{fig:rho_relation}, and $\rho$ and $\rho_i$'s are defined in the caption.
The relation of \eqref{eq:rho_approx_relation} is justified in the Appendix with a proof in the asymptotic case and an observation in the finite-sample case.
With the claimed relationship \eqref{eq:rho_approx_relation}, we investigate the increase in variance after one cut.
We consider the correlation values $\{\rho_i\}_{i=1}^4$ as random variables that are identically distributed.
In the unmatched scenario, the correlation values should have a zero mean and correlation values produced by neighboring blocks that do not have reasons to be dependent.
We used experimental results to confirm that $\cov(\rho_i, \rho_{i^\prime}) = 0, \forall i \neq i^\prime,$ for the unmatched case.
After cutting the heightmap into four subblocks, we calculated correlation values $\{\rho_i\}_{i=1}^4$. There were $81$ correlation values for the $i$th block location, and we ordered them into a vector $\rhov_i$. We used the sample correlation value $\corr(\rhov_i, \rhov_{i^\prime})$ to estimate the theoretical quantity $\corr(\rho_i, \rho_{i^\prime})$. The sample mean and standard deviation values of correlation values $\corr(\rhov_i, \rhov_{i^\prime})$ for subbands \#2--\#4 are around $-0.1$ and $0.2$, respectively.
A $t$-test shows that the correlation values are not significantly different from zero ($p$-value $= 0.249$), which supports our hypothesis.
Hence, by applying the variance operation to \eqref{eq:rho_approx_relation} and using $\cov(\rho_i, \rho_{i^\prime}) = 0$, we obtain for the unmatched scenario:
\begin{equation}
\var(\rho_1) = 4\var(\rho).
\end{equation}
Therefore, after one cut the standard deviation of the correlation values will increase by a factor of $2$, which is consistent with the aforementioned empirical observation.
For the matched case, the correlation values produced by neighboring blocks should be positively correlated, i.e., $\cov(\rho_i, \rho_{i^\prime}) > 0$ for $i \neq i^\prime$. For example, $\{\rho_i\}_{i=1}^4$ are likely to be simultaneously all high or all low, but it is less likely to have two high values and two low values.
We calculated the sample correlation value $\corr(\rhov_i, \rhov_{i^\prime})$. The sample mean and standard deviation values of correlation values $\corr(\rhov_i, \rhov_{i^\prime})$ for subbands \#2--\#4 are around $0.4$ and $0.2$. 
A $t$-test shows that the correlation values are significantly larger than zero ($p$-value $= 8.48\times10^{-8}$), which also supports our hypothesis.
Applying the variance operation to \eqref{eq:rho_approx_relation}  and considering the positive correlation among $\rho_{i} s$, we obtain for the matched scenario:
\begin{equation}
\var(\rho_1) = 4 \var(\rho) - \frac{1}{2} \sum_{i \neq i^\prime} \cov(\rho_i, \rho_{i^\prime}) < 4 \var(\rho),
\label{eq:rho1_matched}
\end{equation}
which corresponds to an increase in standard deviation by a factor of less than $2$ after one cut, which is also consistent with the empirical observation of a factor of $1.5$.

\subsection{Resolution of Norm Map}
Another research question closely related to the issue of the patch size studied in the previous subsection is the choice of resolution for digitizing the patch.
The resolution used in the experiments of this work is $300$ pixels per inch (ppi), or $84.7$ $\mu$m per pixel, i.e., a patch of \PatchDimInch is digitized to $200$-by-$200$ working pixels.
According to Section VII.C and Fig.~14 of [10], within the squared regions of the size of a working pixel, most surfaces ``were not flat because the scale of fibers is smaller than the area of a working pixel.''
Shall we reduce the size of working pixels so that the surfaces corresponding to pixels can be more flat so as to improve the characterization of the structure of the paper, and, in turn, improve the authentication performance?

\begin{figure}[!t]
\centering
  \vspace{-5mm}
  \hspace{-0mm}
  \subfloat[]{\includegraphics[width=0.38\linewidth]{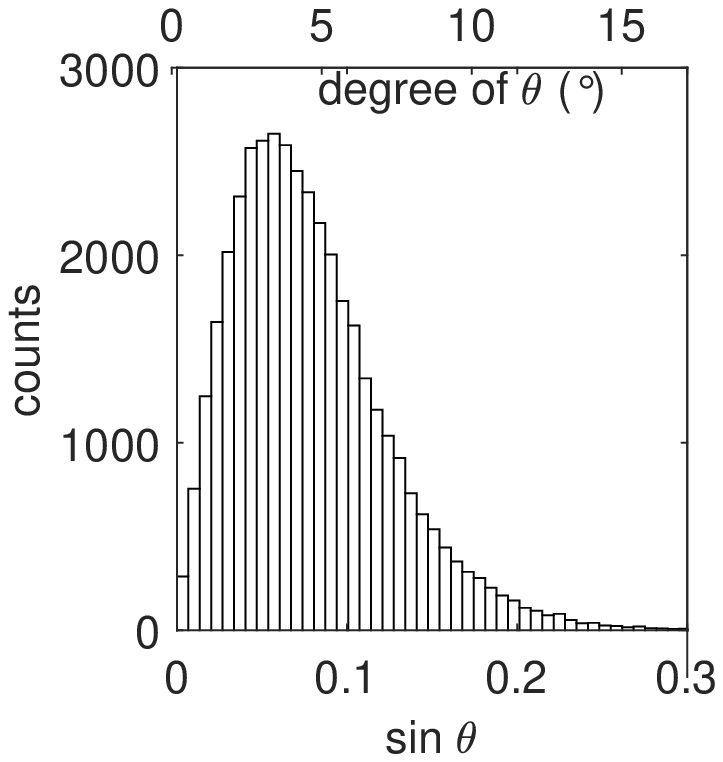}}
	\subfloat[]{\includegraphics[width=0.62\linewidth]{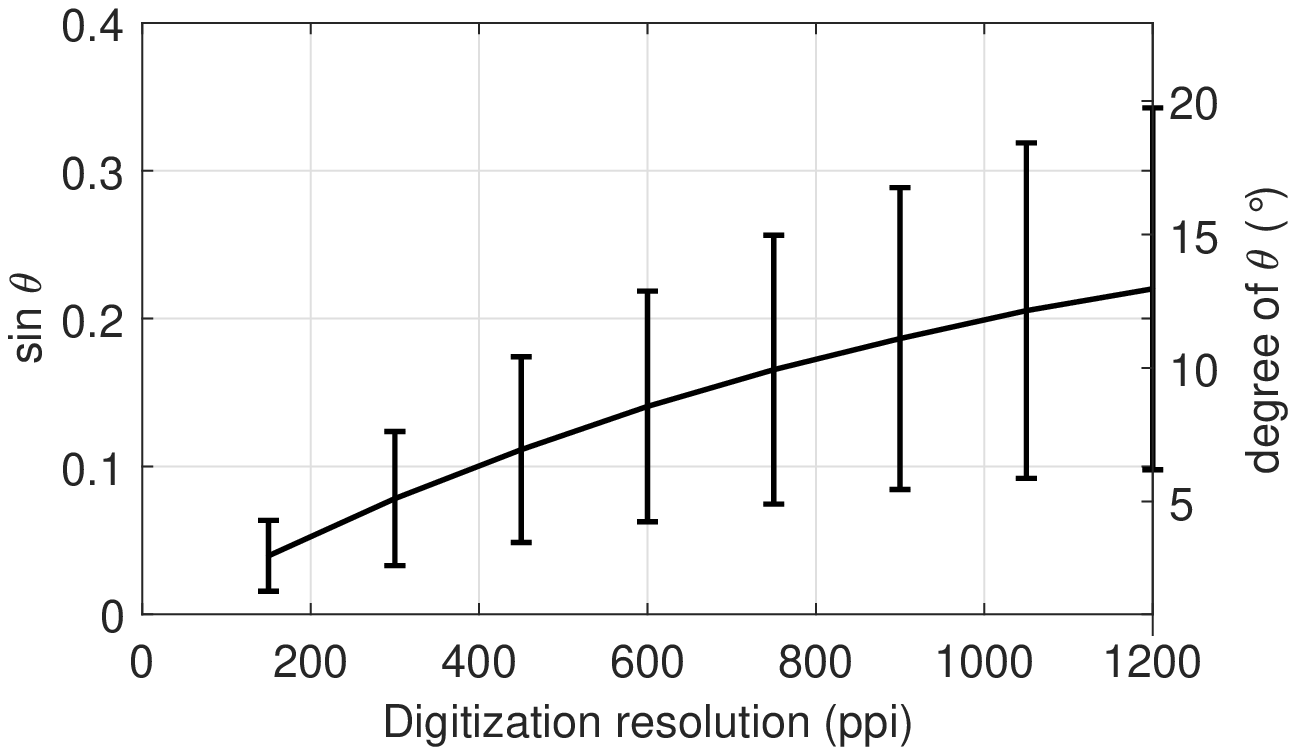}}
  \vspace{-0mm}
  \caption{(a) Histogram of the orientation of squared area covered by a working pixel when a paper patch of size \PatchDimInch is digitized to $200$-by-$200$ working pixels or $300$ ppi. (b) The averaged orientation as a function of digitization resolution. Error bars correspond to one sample standard deviation above and below the average. The monotonic smoothly increasing curve does not 
strongly justify the use of a particular resolution among others within the interior of $[150, 1200]$ ppi.}
  \vspace{-0mm}
  \label{fig:mean_std_theta_squareL}
\end{figure}

We first examine the distribution of the orientations of squared areas of the size of a working pixel when a paper patch of size \PatchDimInch is digitized to $300$ ppi.
We use the tangent plane algorithm in \cite{wong2017} to obtain surface normal vectors using a heightmap captured by a confocal microscope. 
We denote the angle formed by the surface normal vector and $z$-axis by $\theta$. 
A histogram for the sine of the working pixel's orientation, $\sin\theta$, is shown in Fig.~\ref{fig:mean_std_theta_squareL}(a), with a sample mean of $0.078$ (or $4.5^\circ$) and a sample standard deviation of $0.045$ (or $2.6^\circ$) for $\sin\theta$.
These estimated angles are very small compared to the actual angles that could be formed by intertwisted fibers.
However, when considering a relatively larger area covered by a working pixel that may contain multiple fiber segments, it is reasonable that prominently tilted structures are smoothed out. 

Next, we vary the resolution of the norm map obtained from the confocal microscope to see how the distribution of surface orientations may change, and whether there exists any resolution that outperforms others.
We vary the resolution ranging from $150$ to $1200$ ppi to cover a practical working range for consumer-grade flatbed scanners. 
As we increase the resolution, working pixels will shrink in size, leading to larger estimated angles.
At each resolution level, we calculate the sample mean and sample standard deviation of $\sin\theta$ and plot the results in Fig.~\ref{fig:mean_std_theta_squareL}(b).
The plot reveals that both average angle and the angle variation increase as the resolution increases, which is reasonable since finer details of the microstructure of the paper surface are captured.
This means that, by using higher resolution (and a fixed number of pixels), a digitized normal vector field is likely to contain more randomness, and therefore can potentially lead to higher authentication performance by reducing the false negative rate.
However, this monotonic smoothly increasing curve does not strongly justify the use of a particular resolution among others within the interior of the interval ranges from $150$ to $1200$ ppi.
In our proof-of-concept work, we stick to the current digitization resolution, i.e., $300$ ppi, so that the resolution is adequate for authentication while keeping the computational complexity at a reasonable level.

\subsection{Impact of Spatial Registration Error}
\begin{figure}[!t]
\centering
  \vspace{-0mm}
  \hspace{-0mm}
  \includegraphics[width=0.5\linewidth]{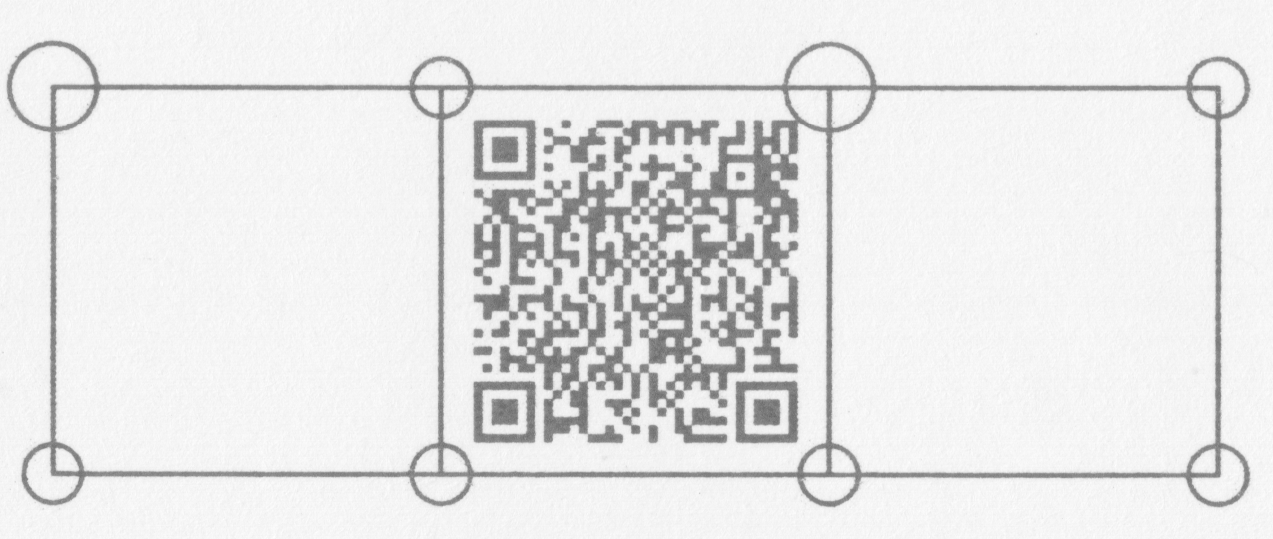}
  \caption{The design of a registration pattern used in this work. The image was captured by a flatbed scanner. The square patch on the left of the QR code is the area used by authentication. By detecting the QR code, the location of the pattern in the image can be roughly estimated, then the precise location is estimated using the lines and circles. Also, the QR code can be used to store information such as paper ID and the reference feature. }
  \vspace{-1mm}
  \label{fig:tripatch}
\end{figure}

In this subsection, we investigate the performance drop due to the error of spatial registration for the paper patch. Clarkson et al. \cite{clarkson-09} applied a lowpass filter to the extracted image and downsampled it to reduce the impact of the registration error, but its effect was not explicitly studied.
Fig.~\ref{fig:tripatch} shows an image of a piece of paper scanned by a flatbed scanner, which shows the design of a registration pattern we used in this work. 
The square patch to the left of the QR code patch is the area of interest that we use for paper surface-based authentication. 
To locate the position of the area of interest, we need to estimate the positions of the intersections. 
We first use a Hough transform to find the lines and then the intersections. We then refine the estimates of the positions of intersections by finding the centers of the circles.
The intersections of the lines in the paper patch are printed in the center of the respective circles.

\begin{figure}[!t]
\centering
  \vspace{-4mm}
	\subfloat[]{\includegraphics[width=0.48\linewidth]{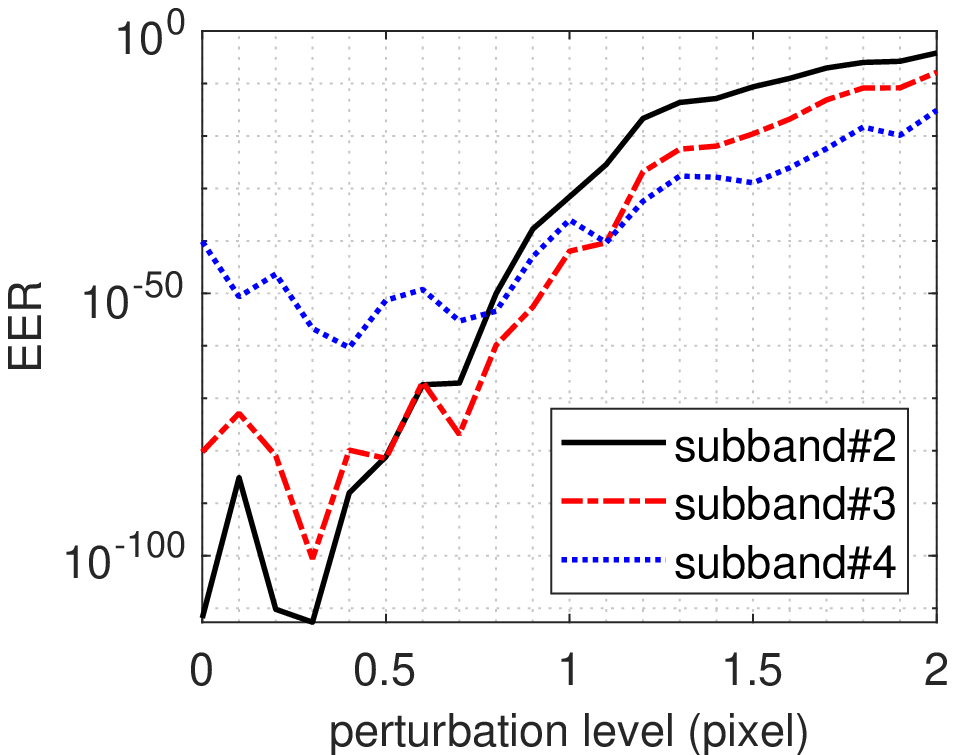}}
	\hspace{2mm}
	\subfloat[]{\includegraphics[width=0.48\linewidth]{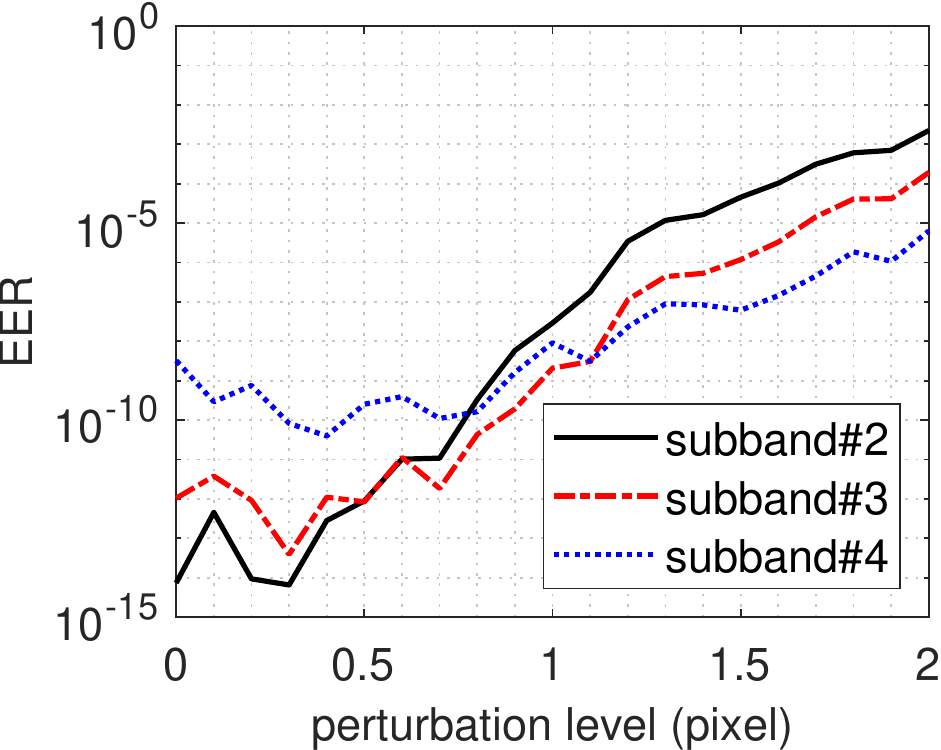}}
  \caption{The impact of spatial registration error: EERs against the perturbation strength $L$ when assuming (a) Gaussian and (b) Laplace distributions. The length of a pixel edge is $\frac{1}{300}$ inches. When there is more registration error (or larger perturbation), the discriminative performance is significantly lowered.}
  \label{fig:eer_perturb}
\end{figure}

In the real-world application, the estimations for the positions of the paper patch may be inaccurate, and, as a result, the performance will drop.
To investigate the effect of imprecise estimations for the positions, we perturb the estimated locations of the four corner positions of the paper patch.
For each of the estimated corner locations $(x, y)$, we add some noise to it, namely, $x' = x + e_1$ , $y' = y + e_2$,  where $e_1, e_2 \sim \mathcal{N}(0, L^2)$, and $L$ is standard deviation to indicate the level of perturbation strength.
We follow the procedure in the practical authentication system in Section \ref{sec:practical} while adding  perturbations to the estimated corner positions in each scanned image of the paper patch.
We increase the perturbation strength $L$ and calculate the EER at each perturbation strength level, and plot the results in Fig.~\ref{fig:eer_perturb}.
When the perturbation strength is small, within $0.3$ pixels, the EERs do not change much. This may be due to the fact that the estimated corner positions of the paper patch were not very accurate in the first place, thus adding small perturbations did not result in much of a performance drop. 
As the perturbation strength increased beyond $0.4$ pixels, the EERs will increase significantly, indicating that deploying a precise image alignment algorithm is one important factor to achieving satisfactory authentication performance.

\section{Discussion}
\label{sec:discussion_application}
The paper surface-based authentication systems can be used in real-world applications. In the systems presented in Fig.~\ref{fig:high_level_diagram}, both mobile cameras \cite{wong2017, liu2018enhanced} and flatbed scanners can be used as the image acquisition devices for authentication systems. Mobile cameras are preferred for commercial applications, such as the authentication of drug packages and wine labels. Both client--server model and local model can be used, but the client--server model can be more flexible in terms of being integrated with other customer services and servers can usually be afforded by for-profit companies. When drugs are packaged by the manufacturer, the reference features of the drug packages will be acquired and saved in a server. After purchasing drugs, the customer can use a mobile phone with a pre-installed app to obtain the test features of the package. The test feature will be sent to the server to compare with the saved reference features, and the final authentication results will be sent back to the customer’s phone. 

When it comes to the verification of important documents, such as birth certificates and academic diplomas, flatbed scanners may be preferred by local governments and academic institutions as scanners are widely deployed. It is less prone to human error for the staff to conduct verification tasks using scanners, and scanners can handle the verification of a large volume of cases when properly automated. For these document issuing entities, the local model is a more sustainable choice than the client--server model mainly because of the long life span of the issued documents. Another advantage of the local model is that even after many generations of technological advancement, one can easily create a verification infrastructure with a new image acquisition device and implement the verification algorithm in a new programming language. When an academic diploma is issued, the reference feature can be acquired and will be encoded into a QR code printed on the diploma. When another institution needs to verify it, a flatbed scanner and a pre-installed verification app can be used to obtain test features and to extract reference features from the QR code. The test and reference features will be compared to determine whether the diploma is authentic.

\section{Conclusion and Future Work} \label{sec:conclusion}
In this work, we have shown by analytic derivations that the specular component of light reflection does not play a role in the estimation of the norm map of paper surfaces in the unique optical setup of a flatbed scanner.
We used a larger dataset to confirm that flatbed scanners can capture meaningful physical quantities of paper surfaces, and we investigated the blurring effect due to the scanner.
We have shown that the high frequency subbands of the reconstructed surfaces are better discriminative features than the norm map, which we verified in a practical engineering system that uses flatbed scanners.
We have shown that larger paper patches will yield better authentication performance in EER, and a precise image alignment algorithm is important for achieving satisfactory authentication performance.

The flatbed scanners instead of mobile cameras have been used as the acquisition device for the studies in this work. Although flatbed scanners are less flexible in terms of portability and acquiring images of objects with irregular shapes such as wine bottles, they have a better-controlled experimental setup. This allows easier investigations into paper surface-based authentication, which is hard to achieve when using mobile cameras in designed experiments. The findings in this work using flatbed scanners may give us insights into how to study research questions for the mobile camera-based authentication system, which are more challenging due to the lower signal-to-noise ratios. In future work, we plan to investigate key research questions on using mobile cameras to acquire the microstructure, e.g., how the specular reflection can be taken into consideration to improve the estimation accuracy of the norm map.

\section*{Acknowledgment}
This work was performed in part at the Analytical Instrumentation Facility (AIF) at North Carolina State University, which is supported by the State of North Carolina and the National Science Foundation (award number ECCS-2025064). The AIF is a member of the North Carolina Research Triangle Nanotechnology Network (RTNN), a site in the National Nanotechnology Coordinated Infrastructure (NNCI).

\appendix
\section*{Justification for $\rho \approx 1/4\sum_{i=1}^4\rho_i$}
\label{app:convergence_proof}

We will provide justification for the relation Eq.~(\ref{eq:rho_approx_relation}) between the sample correlation coefficient of a block, $\rho$, and the sample correlation coefficients of its nonoverlapping, equal-sized subblocks, $\{\rho_i\}_{i=1}^4$, namely, $\rho \approx 1/4\sum_{i=1}^4\rho_i$. We will argue in the finite-sample case that the residual $r_n = \rho -\frac{1}{4} \sum_{i=1}^{4} \rho_{i} \approx 0$. We will also prove that in the asymptotic case $|r_n|$ converges to $0$ in probability.

We denote, for $i$th subblock, the raw data $\x_i^r = (x_{i1}, x_{i2}, \dots, x_{in})$, the sample mean $x_{i\cdot} = \frac{1}{n}\sum_{j=1}^n x_{ij}$, and the mean-removed data $\x_i = \x_i^r - x_{i\cdot}$, where $i \in \{1,2,3,4\}$. 
The mean-removed data for the parent block can be represented as follows:
\begin{equation}
\x
\stackrel{(a)}{=} 
\begin{bmatrix}
\x_1^r \\
\vdots \\
\x_4^r 
\end{bmatrix} - \frac{1}{4}\sum_{i=1}^4 x_{i\cdot}
\stackrel{(b)}{=}
\begin{bmatrix}
\x_1 \\
\vdots \\
\x_4 
\end{bmatrix} + 
\begin{bmatrix}
\epsilon_1 \mathbbm{1} \\
\vdots \\
\epsilon_4 \mathbbm{1} 
\end{bmatrix}
\stackrel{(c)}{=} 
\x^{\prime} + \epsilon,
\label{eq:approx_one}
\end{equation}%
\noindent{}where $\mathbbm{1}$ is length-$n$ vector of all ones, and $ \epsilon_i = \frac{1}{4}(3x_{i\cdot} - \sum_{i^\prime \neq i}^{}x_{i^\prime\cdot})$ is a perturbation term.
Here, (\ref{eq:approx_one}a) and (\ref{eq:approx_one}c) is by definition.
(\ref{eq:approx_one}b) connects the mean-removed terms $\x$ and $\{\x_i\}_{i=1}^4$ at two scales.
With the definitions of $\x$ and $\{\x_i\}_{i=1}^4$, $\rho$ and $\rho_i$ defined in the caption of Fig.~\ref{fig:rho_relation} can be rewritten as:
\begin{equation}
    \rho {=} \frac{\x^T\y}{\|\x\| \|\y\|}, \quad
    \rho_i {=} \frac{\x_i^T\y_i}{\|\x_i\| \|\y_i\|}, \quad i=1,\cdots,4.
\end{equation}

\noindent\textbf{Finite-sample approximation}\hspace{3mm}
For a finite sample size, we justify the following relationship by showing perturbation terms are close to zero and $\|\x_i\|$s are close to $\|\x\|/2$:
\begin{equation}
\begin{split}
r_n &=\sum_{i=1}^{4} \x_{i}^{T}\y_{i}\left(\frac{1}{\|\x\|\|\y\|}-\frac{1}{4\left\|\x_{i}\right\|\left\|\y_{i}\right\|}\right) \\
&+ \left[\mathbbm{1}^{T}\left(\epsilon_{i} \x_{i}+\varepsilon_{i} \y_{i}\right) + n\epsilon_i\varepsilon_i \right] /\|\x\| \|\y\| \approx 0.
\label{eq:finite_sample_approx}
\end{split}
\end{equation}

\noindent Assume that $x_{ij}$'s are independent and identically distributed with mean value $\mu$ and variance $\sigma^2$. 
Note that $\|\x_i\|^2 = \sum_{j=1}^n x_{ij}^2 - n x_{i\cdot}^2$. It is easy to show using the strong law of large number that $\|\x_i\|^2$ converges to $n\sigma^2$ almost surely, and $\|\x\|^2$ and converges to $4n\sigma^2$ almost surely. 
Hence, the term in the parentheses of (19) is close to zero.
Both perturbation terms $\epsilon_i$ and $\varepsilon_i  \sim \mathcal{N}(0,\,0.75\sigma^{2}/n)$ are zero mean with very tiny variance for large $n$, e.g., $n = 10000$ in our application scenario. Hence, the term in the brackets is also close to zero.
We also used real data to verify that $r_n \approx 0$.
We followed the procedures in Section~\ref{subsec:paper_size} to cut the subbands into four subblocks and calculate the sample correlation values $\rho$ and $\rho_i, \ i=1,\cdots,4$. Under the matched case, i.e., the population correlation is larger than zero, the sample mean and standard deviation of $r_n$ for subbands \#2--\#4 were around $10^{-5}$ and $10^{-3}$, respectively.
Under the unmatched case, i.e., the population correlation is zero, the sample mean and standard deviation of $r_n$ for subbands \#2--\#4 were around $10^{-4}$ and $10^{-3}$, respectively.
The small residuals confirmed that $r_n \approx 0$ for the finite-sample scenario.

\noindent\textbf{Lemma 1.}\cite{kenney1951mathematics}\hspace{3mm} When population correlation value $\rho_t$ of a bivariate Gaussian pair is nonzero, the expectation and variance of sample correlation value $\rho$ can be expressed in the form of series:
$\footnotesize
    \E[\rho] = \rho_t-\frac{\rho_t\left(1-\rho_t^{2}\right)}{2 (n-1)}+\cdots,
    \var(\rho) = \frac{\left(1-\rho_t^{2}\right)^{2}}{n-1}\bigg[1+\frac{11 \rho_t^{2}}{2 (n-1)}+\cdots\bigg],
$
where $n$ is the sample size.

\vspace{1mm}
\noindent\textbf{Convergence in mean}\hspace{3mm}
Denote the population correlation value to be $\rho_t$. The sample size is $4n$ for the block and $n$ for a subblock. From Lemma 1, we have:
$
    \E\left[\rho-\frac{1}{4} \sum_{i=1}^4 \rho_{i}\right] 
    =\left(\rho_{t} - \frac{\rho_{t}\left(1-\rho_{t}^{2}\right)}{2 \cdot(4n-1)}+
    \cdots\right)-
    \frac{1}{4}\sum_{i=1}^4\left(\rho_{t}-\frac{\rho_{t}\left(1-\rho_{t}^{2}\right)}{2(n-1)}+\cdots\right)
    \rightarrow 0
$
as $n \rightarrow \infty$.

\vspace{1mm}
\noindent\textbf{Convergence in probability}\hspace{3mm}
For a sample correlation $\rho$ in a block, from Lemma 1 and Markov's inequality we can derive:
\begin{equation}
\begin{split}
    &\mathbbm{P}(|\rho-\rho_{t}|>\varepsilon)\!\leq \! 
         \left(\var(\rho)\!+\!(\E(\rho)\!-\!\rho_{t})^{2}\right)/{\varepsilon^{2}}=
         1/\varepsilon^{2} \\
         & \! \! \cdot  \left[ \tfrac{\left(1-\rho_{t}^{2}\right)^{2}}{4n - 1}\left(1 \! +\! \tfrac{11\rho_t^{2}}{2 (4n-1)} \!+\! \cdots\right) \!+\! \left(\tfrac{\rho_{t}\left(1-\rho_{t}^{2}\right)}{2 (4n-1)}\!+\!\cdots\right)^2\right]\! \! ,
\end{split}
\end{equation}
\noindent which is easy to show that $\mathbbm{P}[|\rho-\rho_{t}|>\varepsilon] \rightarrow 0$ and hence $\rho$ converges to $\rho_t$ in probability, or in a slightly different form $|\rho - \rho_{\text{t}}| \xrightarrow{\text{p}} 0$. Similarly, $|\rho_i - \rho_{\text{t}}| \xrightarrow{\text{p}} 0$.
From triangle inequality, we have 
$|\rho_i - \rho| \leq |\rho_i - \rho_{\text{t}}| + |\rho_{\text{t}} - \rho| \xrightarrow{\text{p}} 0$.
Applying triangle inequality again, we conclude the proof: 
\begin{equation}
	|r_n| = \frac{1}{4}\left|\sum_{i=1}^4 (\rho_i - \rho)\right| \leq \frac{1}{4}\sum_{i=1}^4 |\rho_i - \rho| \xrightarrow{\text{p}} 0.
\end{equation}

\small
\bibliographystyle{IEEEtran}
\bibliography{paper_puf_scanner}

\normalsize
\balance
\vspace{-0mm}
\begin{IEEEbiography}[{\includegraphics[width=1in,height=1.25in,clip,keepaspectratio]{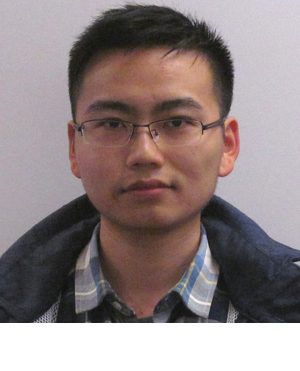}}]{Runze Liu}
(S'18) received his B.E.  in electronic information science and technology from Tsinghua University, Beijing, China, in 2015. He is currently pursuing the Ph.D. degree with the Department of Electrical and Computer Engineering at North Carolina State University, USA. His research interests include machine learning, statistical signal processing, and multimedia forensics.
\end{IEEEbiography}

\vspace{-0mm}
\begin{IEEEbiography}[{\includegraphics[width=1in,height=1.25in,clip,keepaspectratio]{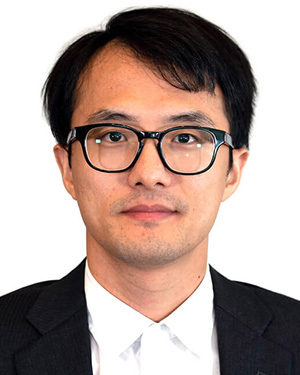}}]{Chau-Wai Wong}
(S'05--M'16) received his B.Eng. and M.Phil. degrees in electronic and information engineering from The Hong Kong Polytechnic University in 2008 and 2010, and the Ph.D. degree in electrical engineering from the University of Maryland, College Park in 2017. He is currently an Assistant Professor at the Department of Electrical and Computer Engineering and the Forensic Sciences Cluster, North Carolina State University. He was a data scientist at Origin Wireless, Inc., Greenbelt, Maryland. His research interests include multimedia forensics, statistical signal processing, machine learning, data analytics, and video coding. Dr. Wong received a Top-Four Student Paper Award, Future Faculty Fellowship, HSBC Scholarship, and Hitachi Scholarship. He was the General Secretary of the IEEE PolyU Student Branch from 2006 to 2007. He was involved in organizing the third edition of the IEEE Signal Processing Cup in 2016 on electric network frequency forensics.
\end{IEEEbiography}

\clearpage
\nobalance
\normalsize
\pagenumbering{gobble}
\placetextbox{0.495}{0.97}{\fbox{\parbox{7in}{\textbf{Note: This is a supplementary document for ``On Microstructure Estimation Using Flatbed Scanners for Paper Surface-Based Authentication,'' published in \emph{IEEE Transactions on Information Forensics and Security} by Runze Liu and Chau-Wai Wong.}
}}}
\vspace*{4mm}

\section*{Supplementary Document}
\setcounter{equation}{0}
\setcounter{figure}{0}
\subsection{Reconstructed Heightmap Leads to Higher Correlation}
\label{app:recon_heightmap}

We follow \cite{liu2018enhanced} to reconstruct heightmaps (3D surfaces) with normal vector fields generated from scanners and a confocal microscope using shapelets \cite{reconstruct} that can be considered as a robust integration algorithm.
The diagrams for generating the heightmaps (3D surfaces) for test and reference patches are shown in Figs.~\ref{fig:diagram}(a) and (b) of the main paper, respectively, excluding the last blocks. The images for the test patch are acquired by scanners and the heightmap for the reference patch is measured by a confocal microscope.
We correlated the reconstructed heightmaps between scanner and confocal microscope, obtaining the correlation at $0.358$ as shown in Table~\ref{tab:corr_summary} of the main paper, which is higher than the correlation at $0.357$ or $0.301$ using the norm map as the feature. 
The improved correlation values indicate that the heightmap with integrated information in both $x$- and $y$-directions is a better discriminative feature than the norm map.

\begin{figure}[!t]
\centering
  \vspace{12mm}
  \hspace{-2mm}
  \subfloat[]{\includegraphics[width=0.48\linewidth]{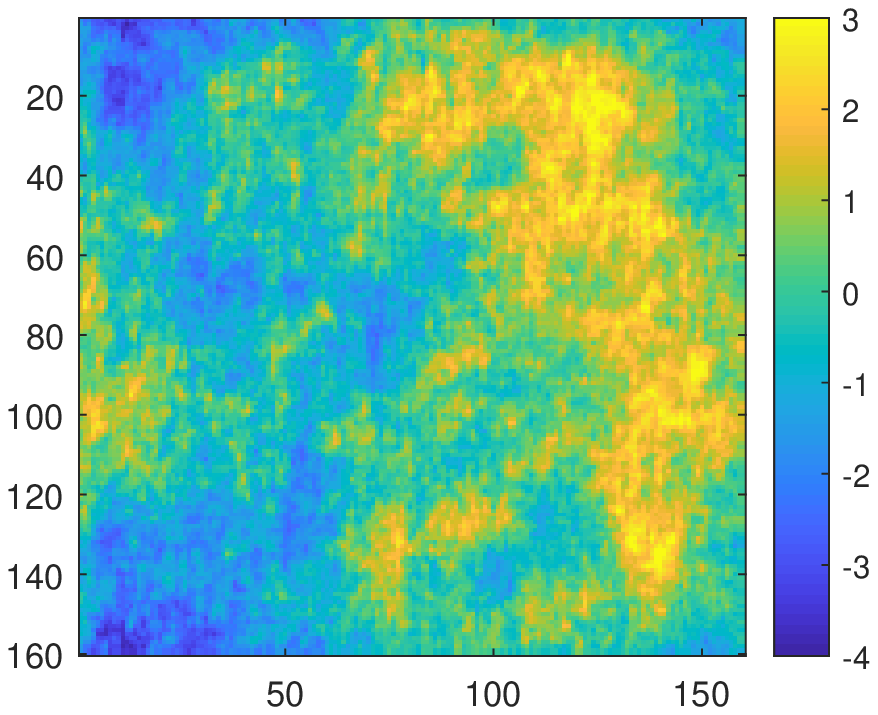}}
  \hspace{2mm}
  \subfloat[]{\includegraphics[width=0.48\linewidth]{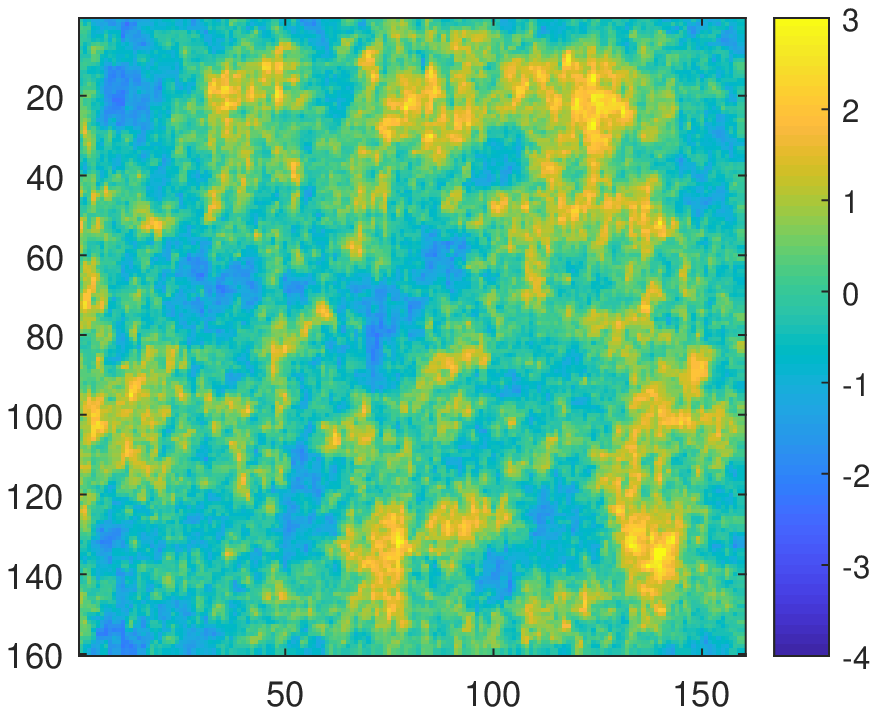}}
  \caption{(a) Reconstructed heightmap from a norm map estimated from images acquired by a scanner, and (b) a detrended version of (a). The detrended heightmap is more flat, and local peaks and valleys are more visible.}
  \vspace{-3mm}
  \label{fig:detrend}
\end{figure}

Fig.~\ref{fig:detrend}(a) shows a reconstructed heightmap from images acquired by a scanner. It is observed that the right part of the paper patch has a higher elevation than the left part. 
This may be caused by the nonflat shape of the paper when scanned, which is not a stable characteristic and may change every time the paper is handled.
The global trend due to the nonflat shape is also problematic from the perspective of the similarity measure using a correlation coefficient: 
i) if two surfaces have similar trends, the correlation between the two surfaces will be high even if their local structures are very different;
ii) if trends are different, the correlation will be low even if their local structures are similar.
Hence, the trend of the heightmaps must be removed before the correlation is calculated.

We removed the trend of the heightmap in Fig.~\ref{fig:detrend}(a) and a detrended version is shown in Fig.~\ref{fig:detrend}(b).
The detrending process contains two steps.
First, a Gaussian blur was applied to generate a surface capturing the overall trend of the heightmap but not capturing the local structures.
In the experiments of this paper in which \PatchDimInch patches are digitized to 200-by-200 pixels, a standard deviation of $25$ pixels was a reasonable value.
Second, the trend surface was subtracted from the heightmap to generate the detrended heightmap.

The correlation resulting from using the detrended heightmap is $0.499$, which is a further improvement over $0.358$, resulting from using the raw heightmap. This result is consistent with that reported in \cite{liu2018enhanced} that studied cameras as acquisition devices. 
Note that the detrended surface retains the middle to high spatial-frequency contents of the raw heightmap that corresponds to local structures, since the trend surface containing the low frequency contents was removed.

\subsection{Discrimination Using Subbands of Heightmap}
\begin{figure*}[thb]
\centering
  \vspace{-6mm}
  \hspace{-1mm}
  \subfloat[]{\includegraphics[width=0.48\linewidth]{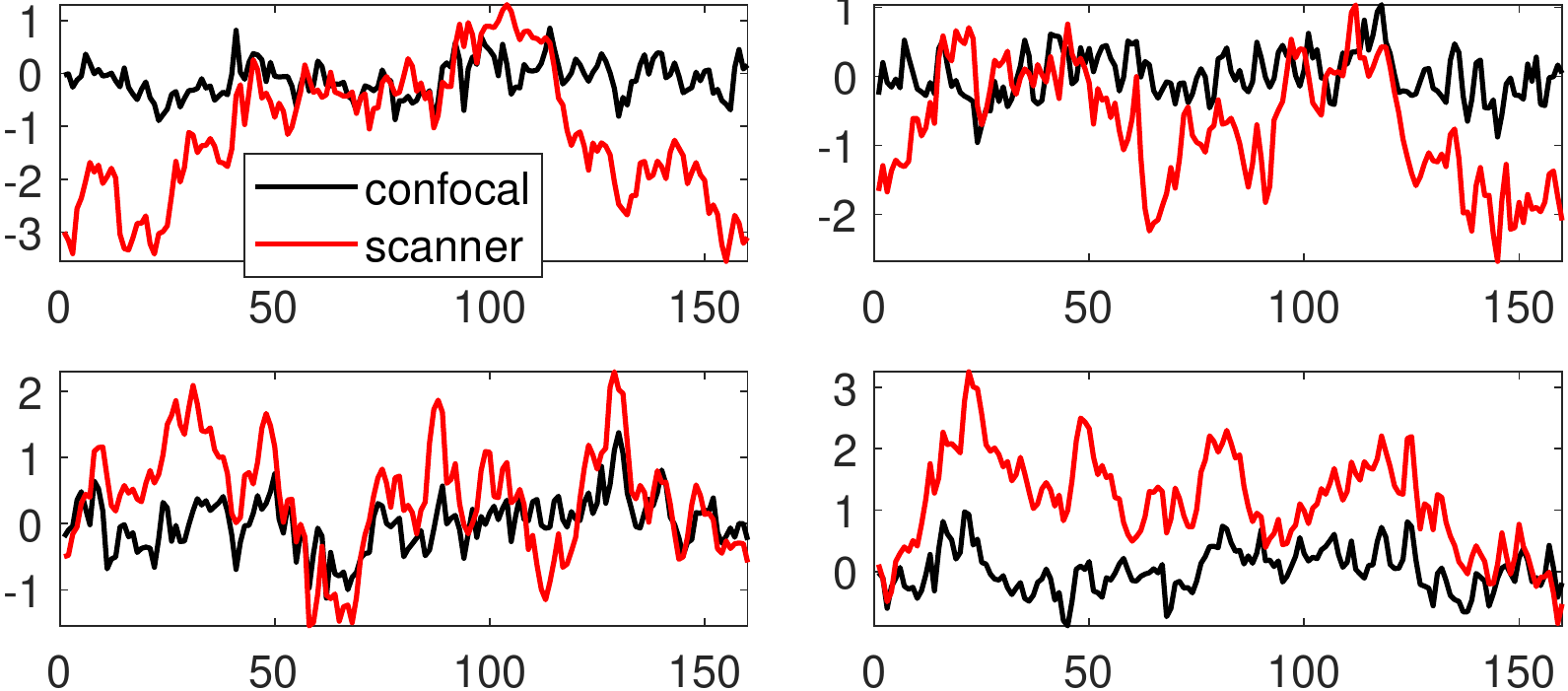}}
	\hspace{4mm}
	\subfloat[]{\includegraphics[width=0.49\linewidth]{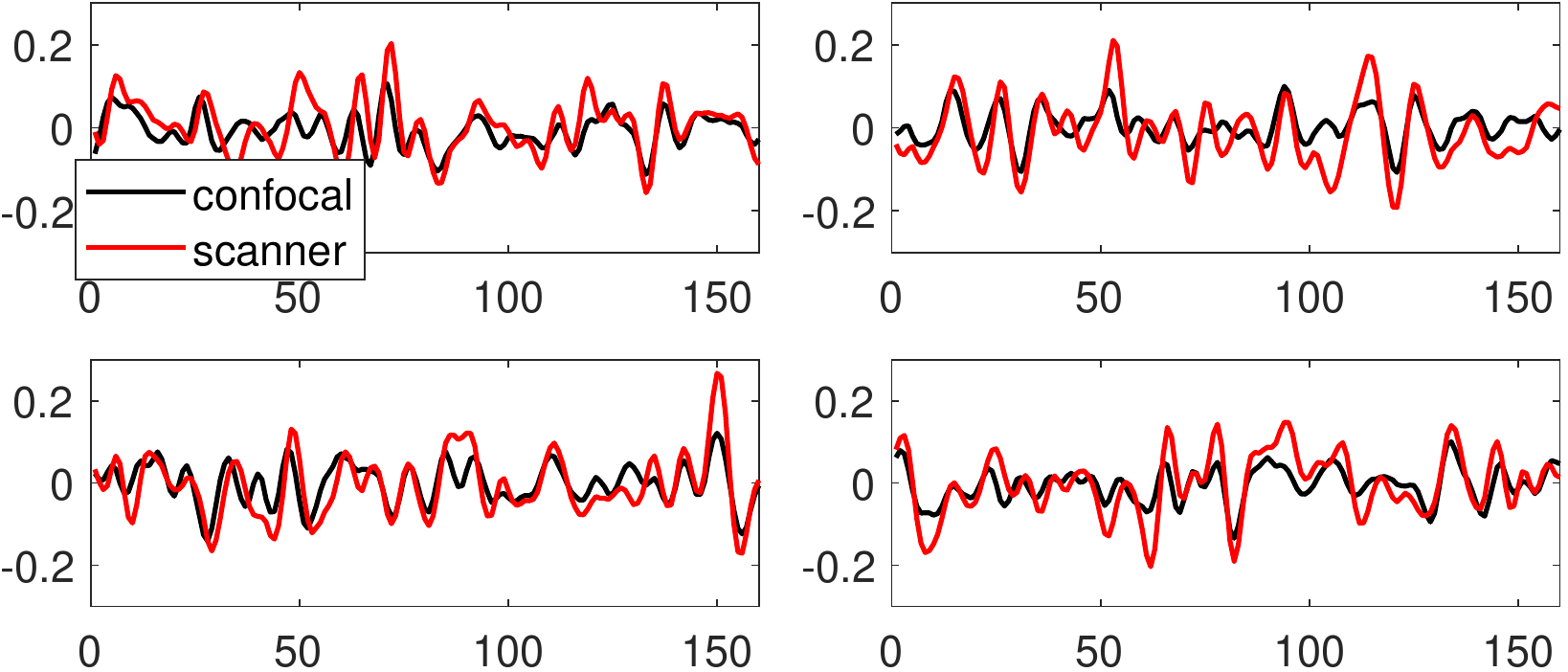}}
  \vspace{-0mm}
  \caption{Representative slices in the $x$-direction from (a) original heightmap and (b) Subband \#3. The slices in the heightmaps of the scanner have trends. The peaks in the high spatial-frequency subbands overlap much better than in the original heightmaps.}
  \label{fig:slices}
\end{figure*}

\begin{figure*}[thb]
\centering
  \vspace{-0mm}
 	\includegraphics[width=\linewidth]{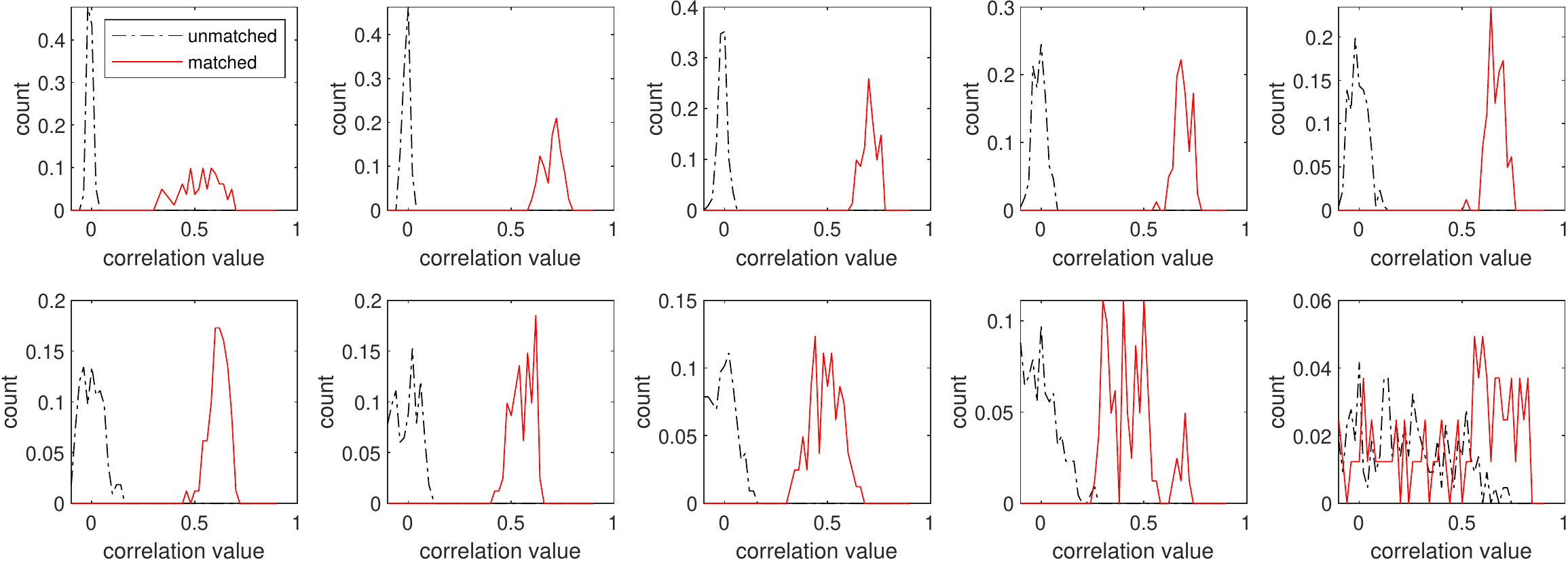}
  \caption{Distributions of correlation values for matched cases and unmatched cases at different subbands. 
    The second and third-highest spatial-frequency subbands are more powerful in describing the uniqueness of physical surfaces.}
  \label{fig:corr_dist}
\end{figure*}
\label{app:performance_heightmap_subband}
The diagrams for generating the subbands of heightmaps (3D surfaces) for test and reference patches are shown in Figs.~\ref{fig:diagram}(a) and (b) of the main paper, respectively, including the last blocks.
We decompose the reconstructed heightmap into ten spatial subbands corresponding to a DoG representation. 
We plot representative slices of the original heightmap in Fig.~\ref{fig:slices}(a) and the third-highest subband, i.e., Subband \#3, in Fig.~\ref{fig:slices}(b).

Fig.~\ref{fig:slices}(a) reveals the trends in the reconstructed surface from scanners. 
Fig.~\ref{fig:slices}(b) shows that the high spatial-frequency subbands from the scanner and confocal microscope match well with each other. 
We calculated the correlation when the scanner matches ($H_0$) or does not match ($H_1$) the confocal microscope for every subband. 
The distributions of correlation values for each subband is shown in Fig.~\ref{fig:corr_dist}.
The distances of the distributions for matched and unmatched in high spatial-frequency subbands are far, indicating a good discriminative capability.
The averaged correlation for best performing subband, i.e., the third-highest spatial-frequency subband, is $0.714$, as shown in Table~\ref{tab:corr_summary} of the main paper.

We quantitatively evaluated the discriminative performance of each spatial subband of the heightmap.
For the majority of them, i.e., Subbands \#1 to \#8, the empirical distributions for the two hypotheses do not have overlap, as shown in Fig.~\ref{fig:corr_dist}. 
This poses a difficulty in estimating discrimination quantities, such as the probability of false alarm or miss when the threshold used is in the middle of two distributions.
This issue is caused by the fact that the overlapping tails are too tiny.
We follow the procedure laid out in \cite{wong2017} to obtain the maximum likelihood estimator (MLE) of the EER using summary statistic quantities of each hypothesis.
Since the EER is achieved when both false-alarm and miss rates are small and equal, the characteristics of extrapolated tails affect the final result significantly.
Since there are not enough data for determining the behavior of the tails, we use a light-tailed distribution, Gaussian, and a heavy-tailed distribution, Laplacian, to quantify the EER in an optimistic way and a pessimistic way, respectively.
It is not difficult to show that, when correlation is assumed to be Gaussian and Laplacian and using a simple thresholding rule, the EER can be written as
\begin{subequations}
\begin{align}
\EER &= \Phi \big[ (\mu_0 - \mu_1) \big/ (\sigma_0 + \sigma_1) \big], \label{eq:eer_gaussian}\\
\EER &= \frac{1}{2}\exp \big[ (\mu_0 - \mu_1) \cdot \lambda_0\lambda_1 \big/ (\lambda_0 + \lambda_1) \big], \label{eq:eer_laplace}
\end{align}
\end{subequations}
respectively, where $\Phi(\cdot)$ is the cumulative density function for the standard Gaussian distribution, and $\mu_i$ and $\sigma_i$, $i=0,1$ are mean and standard deviation for the $i$th hypotheis.
By the invariance principle, we could substitute MLE estimates for $\mu_i$ and $\sigma_i$ into the above equations to obtain the MLE for the EER. 
 
The estimated EER as a function of subband index is shown in Fig.~\ref{fig:eer} of the main paper.
It is revealed that the third-highest spatial-frequency subband is the most discriminative, achieving an EER at $10^{-36}$ or $10^{-8}$ under the Gaussian or Laplacian tail extrapolation assumption, respectively.
We also compare the performance of subbands of heightmap to that of other physical features, i.e., norm map and detrended heightmap, as shown by horizontal lines in Fig.~\ref{fig:eer} of the main paper.
Their EERs are much worse than using the third-highest spatial-frequency subband.

\balance

\end{document}